\documentclass[jgrga]{AGUTeX}
\usepackage{graphicx}
\usepackage{amsmath}
\usepackage{amssymb}
\setkeys{Gin}{draft=false}

\authorrunninghead{STEVENSON, J. E. H. and PARNELL, C. E.}
\titlerunninghead{II. Nature of the waves and flows}
\authoraddr{Corresponding author: J. E. H. Stevenson, School of Mathematics and Statistics, Mathematical Institute, North Haugh, St Andrews, Fife, KY16 9SS, Scotland (jm686@st-andrews.ac.uk)}
\begin{document}

\title{Spontaneous reconnection at a separator current layer.\\ II. Nature of the waves and flows}
\authors{J. E. H. Stevenson,\altaffilmark{1} and C. E. Parnell\altaffilmark{1}}

\altaffiltext{1}{School of Mathematics and Statistics, Mathematical Institute, North Haugh, St Andrews, Fife, KY16 9SS, Scotland}

\begin{abstract}
Sudden destabilisations of the magnetic field, such as those caused by spontaneous reconnection, will produce waves and/or flows. Here, we investigate the nature of the plasma motions resulting from spontaneous reconnection at a 3D separator.
In order to clearly see these perturbations, we start from a magnetohydrostatic equilibrium containing two oppositely-signed null points joined by a generic separator along which lies a twisted current layer. The nature of the magnetic reconnection initiated in this equilibrium as a result of an anomalous diffusivity is discussed in detail in \cite{Stevenson15_jgra}.
The resulting sudden loss of force balance inevitably generates waves that propagate away from the diffusion region carrying the dissipated current. In their wake a twisting stagnation-flow, in planes perpendicular to the separator, feeds flux back into the original diffusion site (the separator) in order to try to regain equilibrium. This flow drives a phase of slow weak impulsive-bursty reconnection that follows on after the initial fast-reconnection phase.
\end{abstract}
\begin{article}
\section{Introduction}
It has long been recognised that magnetic reconnection generates waves and flows since the magnetic energy released by reconnection not only leads to direct heating, but also accelerates populations of particles and the bulk plasma. Indeed, the simple order-of-magnitude estimates of the plasma behaviour in a steady two-dimensional (2D) magnetohydrodynamic (MHD) reconnection scenario \citep{Parker57} revealed that reconnection outflows can be at the Alfv{\'e}n speed. \citet{Petschek64} recognised that such fast outflows could lead to shocks being created in the outflow regions, and developed a steady 2D MHD reconnection model incorporating shocks producing both fast reconnection, as well as additional heating on top of that due simply to Ohmic dissipation alone. 

Many modifications have been made to these models with numerous more complex 2D reconnection scenarios proposed \citep[see, for example, the reviews by][]{PriestForbes,Biskamp2000}. 
In particular, with the ability to perform large-sale numerical experiments, 2D reconnection is now modelled using a wide range of approaches including MHD, Hall-MHD, multi-fluid, hybrid and particle-in-cell \citep[e.g.][]{Birn01}. The addition of extra physics beyond MHD means that instead of just MHD waves (Alfv{\'e}n, fast magnetoacoustic and slow magnetoacoustic) being found, there are also higher frequency waves such as Whistler waves \citep[e.g.,][]{Drake97,Fujimoto08} and ion/electron cyclotron waves \citep[e.g.,][]{Hoshino98,Arzner01}, which can have a significant effect on characteristics such as the onset time and the rate of reconnection.

Reconnection occurs in many geophysical situations, e.g., solar flares, coronal mass ejections, substorms, interactions between planetary magnetospheres and interplanetary magnetic fields (IMF). Observational evidence of fast flows from reconnection has been detailed for many years \citep{Paschmann79,Sonnerup81,Gosling86,Innes97,Phan00,Oieroset00,Yokoyama01,Ko03,Lin05,Wang07,Nishizuka10,Milligan10,Liu10,Hara11,Takasao12,Savage12,Watanabe12,Cao13}. Here, we are specifically interested in the nature of the waves and flows generated as a result of three-dimensional (3D) reconnection: a topic which has not been widely studied. This may be due to several factors: (i) 3D reconnection has many differences to 2D reconnection and identifying exactly where the reconnection occurs and its nature are much harder to do in 3D than in 2D, also (ii) most models of reconnection are driven and so it is difficult to disentangle the waves generated as a result of the reconnection from the flows driven by the boundary conditions. Due to these difficulties we constrain ourselves here simply to studying the waves and flows within a 3D MHD model generated as a result of reconnection that occurs spontaneously in a magnetohydrostatic (MHS) equilibrium.

In 2D there have only been a few models specifically designed to investigate the MHD waves generated by reconnection \citep{Longcope07,FFP12,FFP12b,LongcopeTarr12}. Here, we briefly discuss the results of a 2D MHD model involving undriven (spontaneous) reconnection occurring in a high-beta plasma, whose approach we follow to investigate the MHD waves generated due to 3D reconnection in a high-beta, MHD scenario.

In order to study the nature of the MHD waves generated from 2D X-point reconnection, \citet{FFP12} 
used the approach of first forming a MHS equilibrium with a current layer about a 2D X-point, before studying the reconnection and associated waves at the null embedded in a high-beta plasma.
In \citet{FFP12}, to trigger reconnection in the current layer, which may arise, for instance, as a result of micro-instabilities, an anomalous diffusivity was introduced. The addition of an anomalous diffusivity term, which acts only where the current is greater than a set value, leads to the current layer (and not the enhanced current along the separatrices) being diffused rapidly. 

Waves, launched from the diffusion site at the fast and slow magnetoacoustic speeds, travel outwards leaving a stagnation flow pattern behind in their wake  \citep{FFP12}. This flow is created because the system tries to restore the equilibrium that has been lost, as a result of the reconnection, by rebuilding the current layer, but this simply drives further reconnection. 
The magnetoacoustic waves carry current away from the current layer and propagate enhancements/deficits of plasma pressure in the outflow/inflow regions. Since the fast and slow magnetoacoustic speeds are very similar in a high-beta plasma the outward propagating waves maintain an elliptical shape, although the major axis of the ellipse switches over time as the speed of the waves is quite different in the inflow and outflow regions.
It was found that most of the reconnection in this high-beta case occurred during an initial rapid diffusion phase (which was followed by a second slow reconnection phase driven by the flows left in the wake of the waves).

An identical experiment was run, but with a surrounding low-beta plasma \citep{FFP12b}, although, at the null itself and in its immediate vicinity the plasma is high-beta, as it must be by the definition of a null. In contrast to the high-beta case, there are distinct differences in the propagation of the magnetoacoustic wave pulses because the fast and slow speeds are distinct in a low-beta plasma. Additionally, in the low-beta case \citep{FFP12b} most of the reconnection occurred in the second phase, as opposed to the first, through an impulsive-bursty reconnection regime. Impulsive-bursty reconnection is not achievable in the high-beta experiment due to the low magnitudes of the forces left in the wake of the propagating waves. These differences in flows are highlighted by the fact that the amplitude of the propagating waves of the high-beta case are $10^{5}$ times smaller than those of the low-beta case.
 


As already briefly mentioned, reconnection in 3D is fundamentally different to that in 2D and it can occur at topological or geometrical features of a magnetic field \citep{Hesse88,Schindler88}. In this paper, we focus on the reconnection which occurs at a topological feature called a separator, since such features have been shown to be prime locations for reconnection \citep[e.g.,][]{Sonnerup79,LauFinn,Longcope96,GN97,G00,Longcope01,PC06,Priest05,Haynes07,Parnell08,Dorelli08,Parnell10a,Komar13,Stevenson15}.

Only generic separators exist for more than an instant in dynamic 3D magnetic fields (separators formed by the intersection of the spines, or one spine and a separatrix surface, from two distinct 3D nulls are non-generic as any small perturbation in the field will destroy the intersection) and are formed by the intersection of the separatrix surfaces of a pair of 3D null points and so are special field lines that connect two null points \citep[see][for a basic discussion on 3D null points, separators and separator reconnection]{LauFinn,Parnell10a,Stevenson15_jgra}. 

Although, it has been established that separators are a common topological feature found throughout the solar atmosphere \citep[e.g.,][]{Close05,Platten14,Parnell15philtrans}, here we focus on those which may exist in the chromosphere, the Earth's magnetosphere or other planetary atmospheres. Notably, in the dayside magnetopause, which is a high-beta plasma region \citep{Trenchi2008}, it is well known that reconnection will most likely occur along the line which separates the four topological flux domains containing the interplanetary magnetic field (IMF), the closed terrestrial magnetospheric field lines and the open field lines that extend from the Earth out into the IMF or extend down from the IMF to the Earth. 
From a 2D perspective, these four flux domains only come together at a single point which must be a null point. In 3D, however, these four domains come together all the way along a line, the field line know as the separator, which crucially does not have zero magnetic field all the way along it. The local magnetic field in planes perpendicular to a separator may be either X-type or O-type in nature, as demonstrated both analytically and numerically by \citet{Parnell10a} and also found in \citet{Stevenson15_jgra}. Thus, the name ``X-line'', which has in the past been used to refer to such a line, is inappropriate and should simply be reserved for scenarios in 2.5D.

On the dayside magnetopause many models have been formulated to predict where the location of this reconnection occurs \citep[e.g.,][]{Sonnerup74,Gonzalez74,Alexeev98,Moore02,Trattner07,Swisdak07,Borovsky08,Dorelli09,Borovsky13,Hesse13}. \citet{Komar13} have mapped the dayside magnetopause separators in global magnetospheric simulations for arbitrary clock angle of the IMF and found that in all cases separators exist and are the locations at which the reconnection occurs.

This paper is the second of a series. In the first paper \citep{Stevenson15_jgra} the nature of the magnetic reconnection which occurred at a single-separator MHS equilibrium current layer, embedded in a high-beta plasma, was studied. In this paper, we focus on the properties of the MHD waves and flows generated as a consequence of separator reconnection. In order to achieve this we study the local region about an isolated straight separator. Obviously, such an idealised scenario is unlikely to be realised in the solar chromosphere, or any planetary magnetosphere. However, our qualitative results should be applicable in any of these scenarios (under the constraints of MHD) and our dimensionless results may be scaled using dimensional factors to produce values that can be compared with results from larger-scale numerical models or observed quantities.

We begin in Sect.~\ref{sec:results} by briefly summarising the properties of the reconnection which are discussed in full in \citet{Stevenson15_jgra} and then recap the details of the initial setup and the MHD code used to carry out the reconnection experiment (Sect.~\ref{sec:equb}). In Sect.~\ref{sec:phase2} we analyse the waves launched, due to the reconnection, and then look at the transport of energy in the system (Sect.~\ref{sec:transp}). Finally, we 
summarise our findings in Sect.~\ref{sec:conclusions}.

\section{Nature of the Reconnection}\label{sec:results}
\citet{Stevenson15_jgra} studied the properties of the reconnection which occurred at a separator current layer embedded in a high-beta plasma. They found that the reconnection occurred in two distinct phase; a fast-reconnection phase ($0t_f \le t \le 0.09t_f$) in which $75\%$ of the magnetic energy was converted into internal and kinetic energy and a slow, impulsive-reconnection phase ($0.09t_f < t \le 0.76t_f$) in which only short-lived sporadic reconnection events occur. All times in the experiment discussed in \citet{Stevenson15_jgra}, and discussed here, are normalised to the time it would take a fast-magnetoacoustic wave to travel along the MHS equilibrium separator from one null to the other 
($t_f=0.88$). The experiment was stopped at $t=0.76t_f$ since the waves, launched from the diffusion site at the start of the reconnection experiment, neared the boundaries at this time. The sporadic reconnection events which occur in phase II were numerous enough such that the total flux reconnected continued to increase during this phase. The reconnection was observed to occur asymmetrically along the entire length of the separator and had a counter-rotating flow associated with it. 

Throughout both phases of the reconnection experiment, Ohmic heating dominated over viscous heating (and adiabatic cooling) due to the model being embedded in a high-beta plasma. The value of the plasma beta, which is high due to the presence of two null points in the system, inhibits the waves from becoming large throughout the experiment. The strongest regions of reconnection occurred along the separator away from the nulls and are associated with elliptical magnetic field lines in planes perpendicular to the separator. The length of the first (main reconnection) phase and the amount of magnetic energy which is converted was shown to be dependent on not only the size of the diffusivity, $\eta_d$, but also the size of the diffusion region, $j_{crit}$.

The sudden loss of force balance in the MHS equilibrium caused by the anomalous diffusivity, leads to waves being launched from the diffusion site which travel outwards and cause changes to both the plasma and the magnetic field in their wake. In this paper, we detail the nature of these waves and flows set up in the system by the reconnection, and analyse how they affect different plasma parameters. In the next section, we briefly outline the properties of the MHS equilibrium current layer which we use as our initial state in our separator reconnection experiment and summarise the numerical model used for completeness.

\section{Initial MHS Equilibrium Current Layer and Numerical Model}\label{sec:equb}
\begin{figure}[h!]
\centerline{
\includegraphics[width=0.45\textwidth,clip]{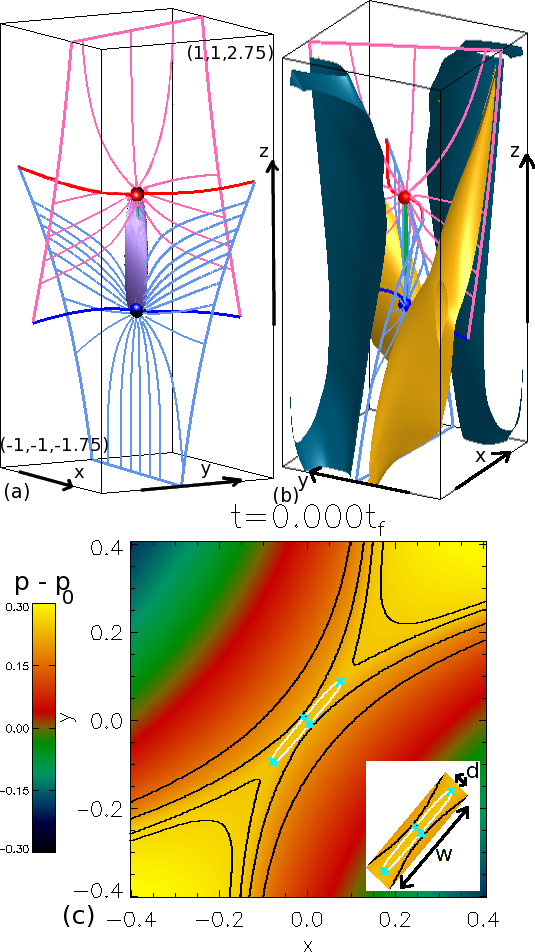}}
\vspace{0.01\textwidth}
\caption{Skeleton of the MHS equilibrium magnetic field with (a) purple isosurface of $j_{\parallel}=10.0$ and (b) yellow/blue isosurfaces of the pressure difference ($p-p_0$, where $p_0=1.5$) drawn at 70\% of the maximum positive/negative values. Also shown are the positive/negative nulls (blue/red spheres) with associated spines (blue/red lines) and separatrix-surface field lines (pale-blue/pink lines) and the separator (green line, hidden by the current layer in (a)). The solid pale-blue/pink lines indicate where the separatrix surfaces intersect the domain boundaries. (c) Perpendicular cut across the MHS equilibrium separator at $z=0.4$ showing contours of the pressure difference with black and white lines showing contours of the magnitude of the current ($|{\bf{j}}|$). Eight cyan asterisks are drawn in four positions on the edge of the white contour  ($j_{crit}=10$) which represents the edge of the diffusion region in this plane. The insert highlights the depth ($d$) and width ($w$) of the diffusion region in this plane.}\label{fig:equbskel}
\end{figure}

The initial state of our reconnection experiment is a MHS equilibrium which contains a twisted 3D current layer lying along the separator. Fig.~\ref{fig:equbskel}a shows the MHS equilibrium skeleton along with the current layer (represented by an isosurface of current drawn at $j_{crit}=10$: the current above which the diffusivity is non-zero). This MHS equilibrium current layer was formed through the non-resistive MHD relaxation of a non-force free magnetic field which contained two null points of opposite signs, whose separatrix surfaces intersected along the $z$-axis to form a generic separator. The full details of how a similar MHS equilibrium was formed and the properties of the separator current layer are given in \citet{Stevenson15}. 

The plasma pressure in this equilibrium is such that pressure enhancements lie in cusp regions about the separator, and the pressure falls off away from here. Fig.~\ref{fig:equbskel}b shows the skeleton of the equilibrium field with yellow/blue isosurfaces of the pressure difference (the pressure in the equilibrium state, $p$, minus the uniform pressure in the domain before the non-resistive relaxation takes place, $p_0=1.5$). Contours of the equilibrium pressure difference are shown in a plane perpendicular to the separator at $z=0.4$ in Fig.~\ref{fig:equbskel}c. Over plotted here are black and white contours of the magnitude of the current $|{\bf{j}}|$ in this plane. The white contour is drawn at $j_{crit}=10$ which is the value that represents the separator current layer in this plane.  Eight cyan asterisks (shown in four positions) are plotted in Fig.~\ref{fig:equbskel}b which lie on the ``edges'' of the MHS equilibrium current layer on the current contour equal to $j_{crit}=10$. The positions of these asterisks will be used (Sect.~\ref{sec:phase2}) to highlight the speed at which waves, launched by the onset of reconnection, move out from the separator current layer. 

This MHS equilibrium is used in \citet{Stevenson15_jgra} as the initial state in a resistive MHD experiment using the Lare3d code \citep{Arber01}. Line-tied boundary conditions are used (the normal components of the magnetic field, density and internal energy per unit mass attain minima or maxima on the boundaries) and the velocity is set to zero on the boundaries. Reconnection is triggered at the separator current layer, which existed in the MHS equilibrium, through the use of an anomalous diffusivity which is zero unless the current is greater than the value, $j_{crit}$, where it takes the value $\eta_d$. As in \citet{Stevenson15_jgra}, we use $j_{crit}=10.0$, such as to include the strong current in the separator current layer in the reconnection, but not the enhanced current on the separatrix surfaces, $\eta_d=0.001$ and a constant background viscosity of $\nu=0.01$. 

Below, we detail the nature of the waves and flows which are created in the system as a result of the spontaneous reconnection discussed in \citet{Stevenson15_jgra}.

\section{Propagation of MHD Waves}\label{sec:phase2}
Initially the plasma is in a MHS equilibrium, but as soon as the current in the current layer starts to dissipate, due to the onset of localised reconnection, waves are launched from the edges of the diffusion region (main current layer Fig.~\ref{fig:equbskel}). These waves travel throughout the system communicating the collapse of the current layer and the resulting loss of force balance. In their wake, the magnetic field and plasma respond to these changes. In this section, we describe the nature of the waves launched and the resulting response of the plasma after they have passed.

In order to investigate these waves we consider the perturbed current, which we define as $|{\bf{j}}|$-$|{\bf{j}}_{MHS}|$, rather than as $|{\bf{j}}-{\bf{j}}_{MHS}|$ such that we can see both enhancements and deficits in the magnitude of the current (top row of Figs.~\ref{fig:perpconts}, \ref{fig:vertcontsd} and \ref{fig:vertcontsw}). We also examine the perturbed pressure, $p$-$p_{MHS}$ (middle row of Figs.~\ref{fig:perpconts}, \ref{fig:vertcontsd} and \ref{fig:vertcontsw}) and the vorticity (bottom row of Figs.~\ref{fig:perpconts}, \ref{fig:vertcontsd} and \ref{fig:vertcontsw}) with snapshots shown at three different times to illustrate their behaviour. The three times which we show represent the experiment near the start of phase I ($t=0.019t_f$), at the end of phase I ($t=0.09t_f$) and about $75\%$ of the way through phase II ($t=0.60t_f$). The waves launched are very small, with amplitudes of order $10^{-3}$  in current and of order $10^{-4}$  in pressure. The amplitudes of the waves are small in comparison to the MHS equilibrium values and reflect the size of the disturbance that caused them, namely the dissipation of the current layer. As discussed in \citet{Stevenson15_jgra}, a high-beta plasma contains current layers that are fatter and have smaller maximum current than the identical low-beta plasma scenario, due to the stabilizing effect of the pressure gradient force in high-beta plasmas. See movies online for the full evolution of the current, pressure and vorticity in the planes shown in Figs.~\ref{fig:perpconts}, \ref{fig:vertcontsd} and \ref{fig:vertcontsw}. 

Cuts in the $z=0.4$ plane are plotted in Fig.~\ref{fig:perpconts} showing the behaviour of the perturbed current, perturbed plasma pressure and vorticity at three different times. In Figs.~\ref{fig:vertcontsd} and \ref{fig:vertcontsw}, the ‘cuts’ are taken at the same three times, but are through the depth and across the width of the current layer, respectively. So, the vertical axis for these graphs runs along the $z$-axis from the lower to the upper null. However, the current layer is twisted and so the horizontal axis changes with distance along the separator, such that it always goes through the current layer depth (Fig.~\ref{fig:vertcontsd}) or width (Fig.~\ref{fig:vertcontsw}), as required. On all these graphs, we have plotted asterisks, which at the start of the experiment are located in the $z=0.4$ plane on the current contour $|{\bf{j}}|=j_{crit}$, i.e., on the edge of the diffusion region (shown in the equilibrium field in Fig.~\ref{fig:equbskel}c). We advect these asterisks through the depth and across the width of current layer in the $z=0.4$ plane at the fast-magnetoacoustic speed, $c_f$, both outwards away from, and inwards across, the current layer, in order to estimate the location, at a given time, of any fast-mode waves launched at the onset of the reconnection. Note, though, since the domain is three dimensional, the waves do not have to move in a planar manner, which may explain the small discrepancies between the wave fronts and the asterisks. 

\subsection{Current and Pressure Perturbations}
\begin{figure*}[h!]
\centerline{
\includegraphics[width=0.065\textwidth,clip]{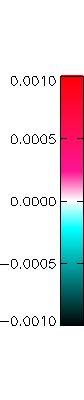}
\includegraphics[width=0.3\textwidth,clip]{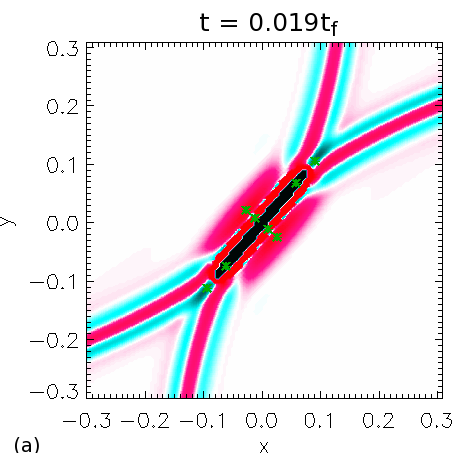}
\includegraphics[width=0.3\textwidth,clip]{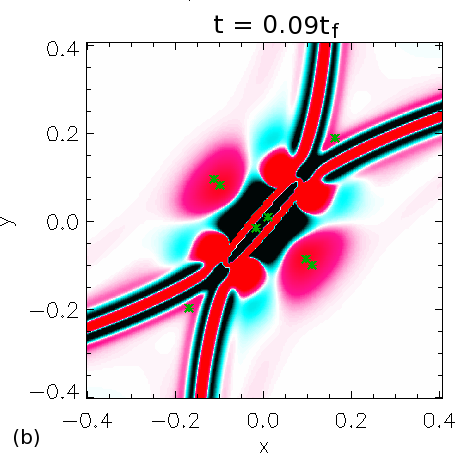}
\includegraphics[width=0.295\textwidth,clip]{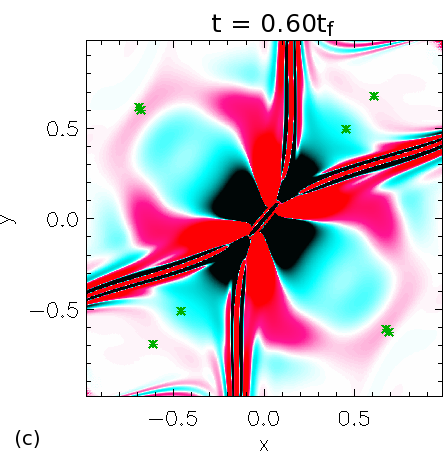}}
\centerline{
\includegraphics[width=0.065\textwidth,clip]{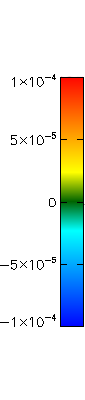}
\includegraphics[width=0.3\textwidth,clip]{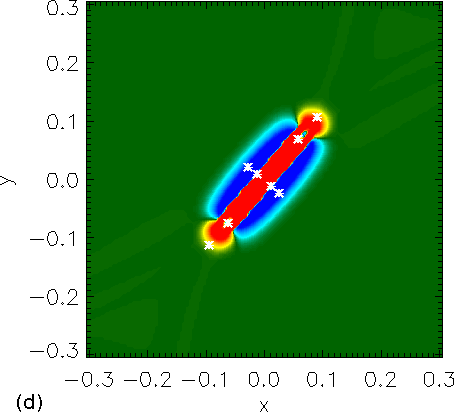}
\includegraphics[width=0.3\textwidth,clip]{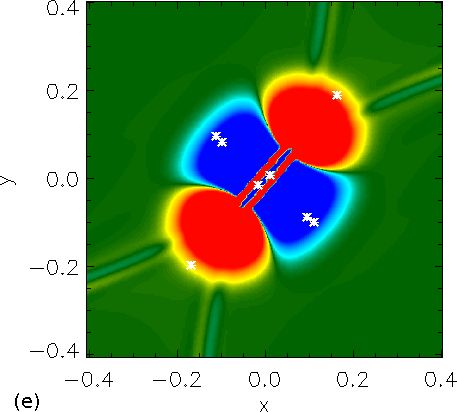}
\includegraphics[width=0.295\textwidth,clip]{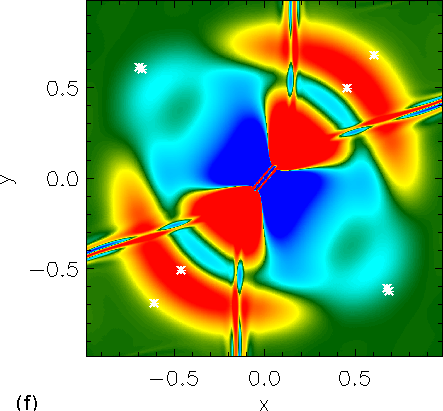}}
\centerline{
\includegraphics[width=0.065\textwidth,clip]{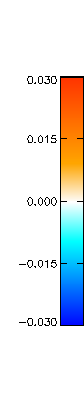}
\includegraphics[width=0.3\textwidth,clip]{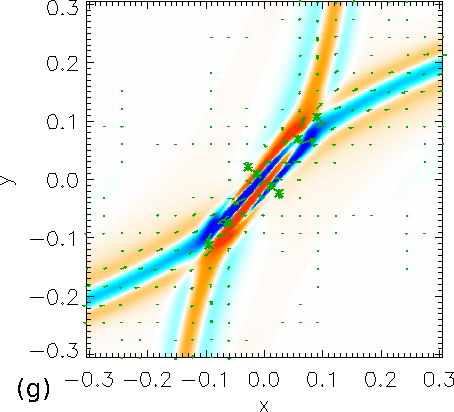}
\includegraphics[width=0.3\textwidth,clip]{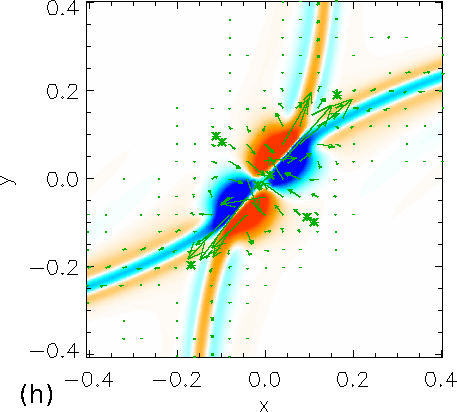}
\includegraphics[width=0.295\textwidth,clip]{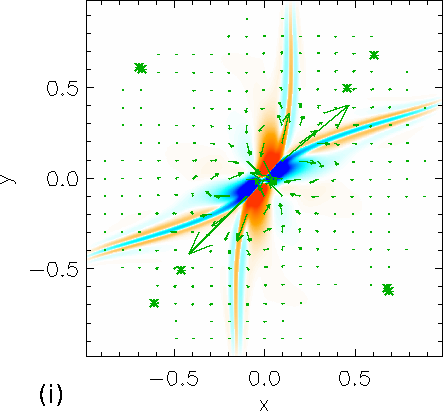}}
\caption{Contours of $|{\textbf{j}}|-|{\textbf{j}}_{MHS}|$ (first row), $p-p_{MHS}$ (second row) and $\omega_z$ (third row) in the plane $z=0.4$ across the separator, at $t=0.019t_f$ (first column), $t=0.09t_f$ (middle column) and $t=0.60t_f$ (last column). Asterisks, which initially lie on the edge of the diffusion region, as shown in Fig.~\ref{fig:equbskel}c, move at the fast-magnetoacoustic speed, $c_f(x,y,z,t)$. Over plotted on the bottom row of graphs are arrows (normalised to the maximum value of the magnitude of the velocity in the domain at $t=0.60t_f$, $|{\bf{v}}|=6 \times 10^{-3}$) that display the direction of $v_x$ and $v_y$ in $z=0.4$ plane. As time increases, so do the dimensions of the planes.}\label{fig:perpconts}
\end{figure*}

\begin{figure*}[h!]
\centerline{
\includegraphics[width=0.07\textwidth,clip]{colbar_diffj_grid_v2.png}
\includegraphics[width=0.3\textwidth,clip]{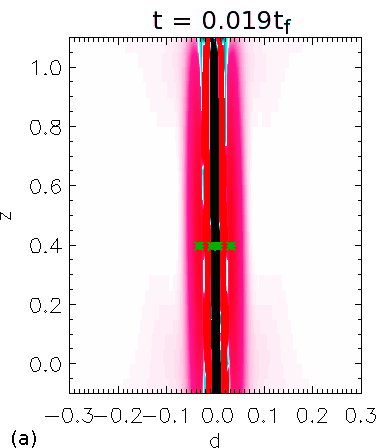}
\includegraphics[width=0.3\textwidth,clip]{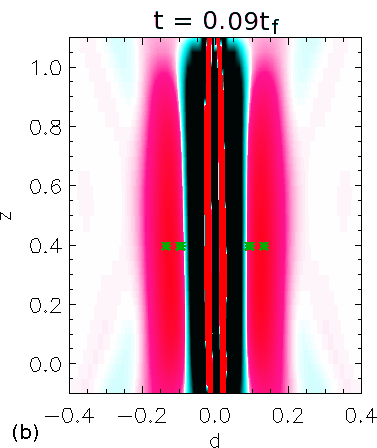}
\includegraphics[width=0.3\textwidth,clip]{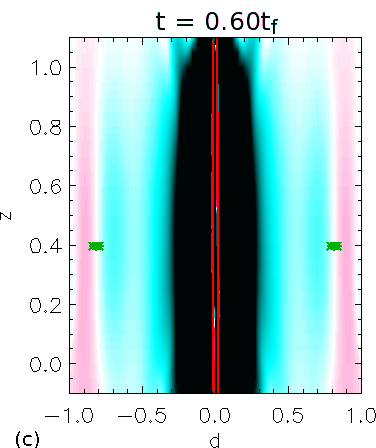}}
\centerline{
\includegraphics[width=0.07\textwidth,clip]{colbar_pdiff_grid.png}
\includegraphics[width=0.3\textwidth,clip]{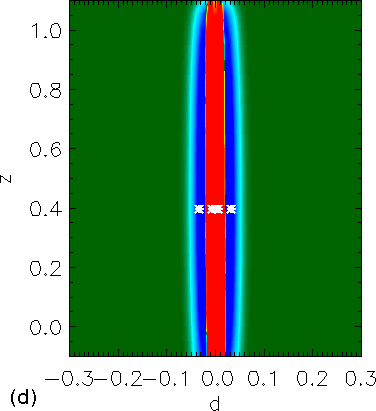}
\includegraphics[width=0.3\textwidth,clip]{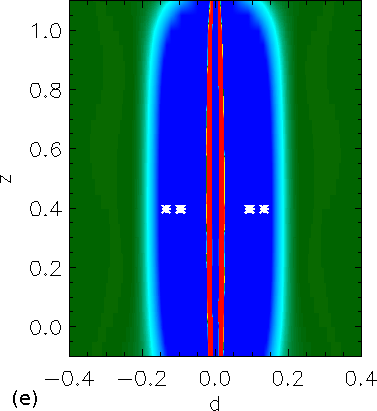}
\includegraphics[width=0.3\textwidth,clip]{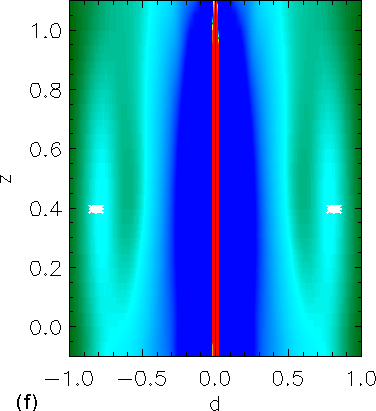}}
\centerline{
\includegraphics[width=0.07\textwidth,clip]{colbar_curlv_grid2.png}
\includegraphics[width=0.3\textwidth,clip]{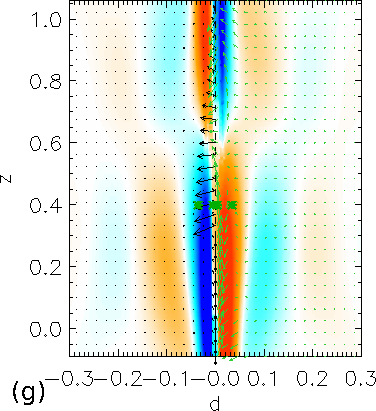}
\includegraphics[width=0.3\textwidth,clip]{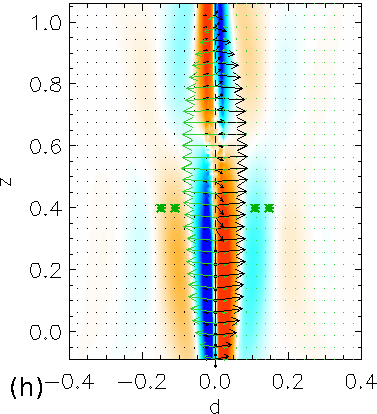}
\includegraphics[width=0.3\textwidth,clip]{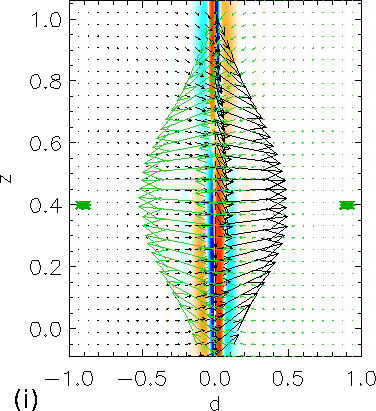}}
\caption{As for Fig.~\ref{fig:perpconts}, but instead showing the perturbations in
a vertical surface that crosses the depth of the current layer at right angles to its width.
Here, the arrows (normalised to the maximum value of the magnitude of the velocity in the domain at $t=0.60t_f$, $|{\bf{v}}|=6 \times 10^{-3}$) display the direction of $v_r$ and $v_z$ in this plane in the last row. The arrows are coloured black where $d \textless 0$ and are coloured green where $d \textgreater 0$. 
}\label{fig:vertcontsd}
\end{figure*}

\begin{figure*}[h!]
\centerline{
\includegraphics[width=0.07\textwidth,clip]{colbar_diffj_grid_v2.png}
\includegraphics[width=0.3\textwidth,clip]{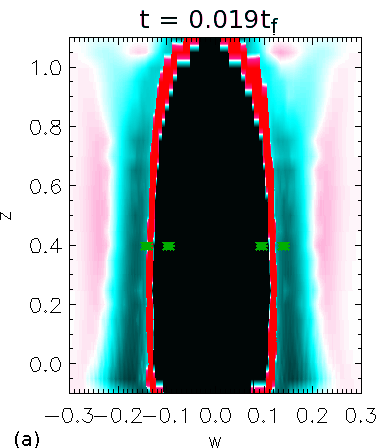}
\includegraphics[width=0.3\textwidth,clip]{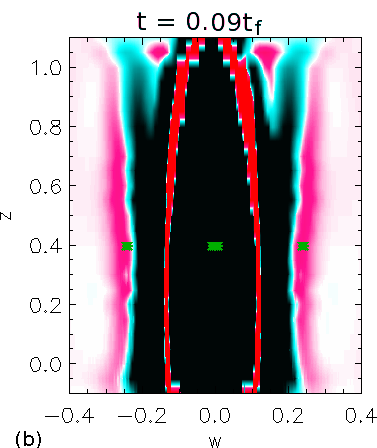}
\includegraphics[width=0.3\textwidth,clip]{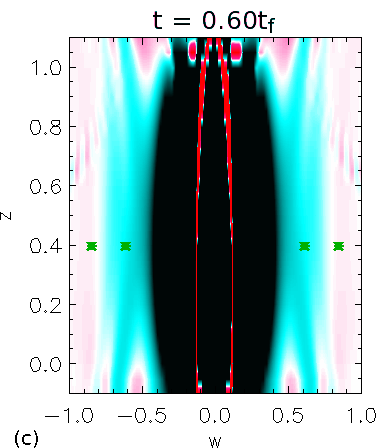}}
\centerline{
\includegraphics[width=0.07\textwidth,clip]{colbar_pdiff_grid.png}
\includegraphics[width=0.3\textwidth,clip]{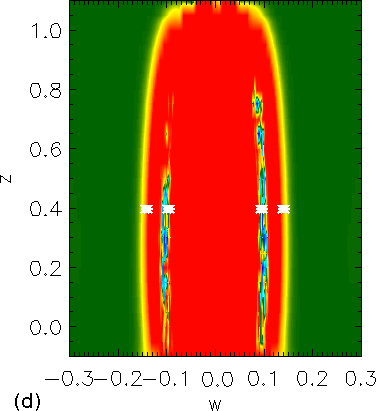}
\includegraphics[width=0.3\textwidth,clip]{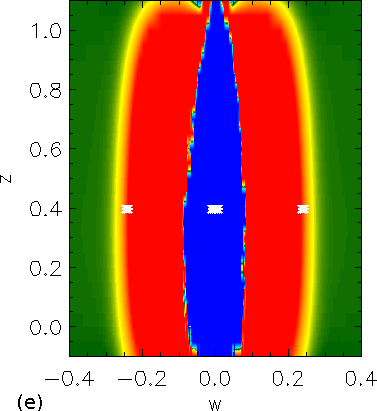}
\includegraphics[width=0.3\textwidth,clip]{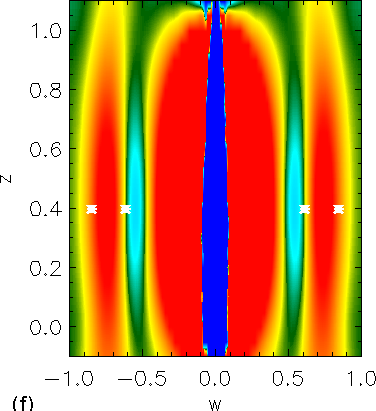}}
\centerline{
\includegraphics[width=0.07\textwidth,clip]{colbar_curlv_grid2.png}
\includegraphics[width=0.3\textwidth,clip]{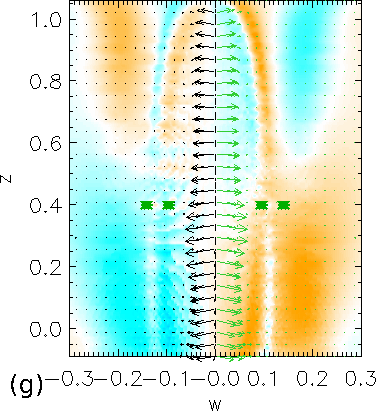}
\includegraphics[width=0.3\textwidth,clip]{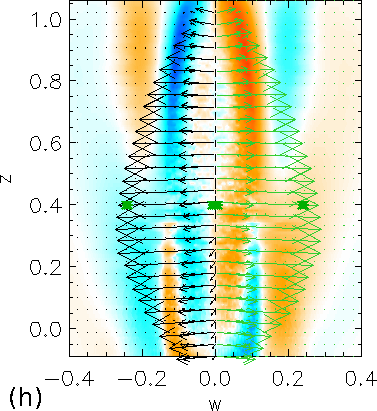}
\includegraphics[width=0.3\textwidth,clip]{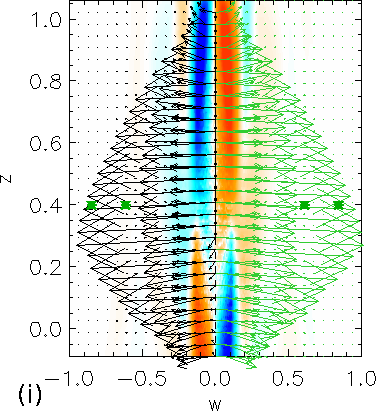}}
\caption{As for Fig.~\ref{fig:vertcontsd}, but instead showing the perturbations in
a vertical surface that crosses the width of the current layer at right angles to its depth.
Here, the arrows are coloured black where $w \textless 0$ and are coloured green where $w \textgreater 0$ and are normalised as in Fig.~\ref{fig:vertcontsd}. As in the previous figures, as time increases, so do the dimensions of the planes.}\label{fig:vertcontsw}
\end{figure*}

The onset of reconnection causes a sudden deficit in current at the current layer, but the current does not disappear, instead it is carried away from the reconnection site by wave pulses that travel at the fast and slow magnetoacoustic wave speeds. Since the plasma is high beta the fast and slow magnetoacoustic wave speeds are very similar in our experiment. Shortly after these waves have been launched (left-hand column of Figs.~\ref{fig:perpconts}, \ref{fig:vertcontsd} and \ref{fig:vertcontsw} at $t=0.019t_f$), a distinctive wave pattern is setup in perturbed current and pressure. In the plane perpendicular to the separator (Fig.~\ref{fig:perpconts}) this pattern is essentially the same as that found following the onset of reconnection at a 2D current layer \citep{FF11}. The leading peaks of the current perturbations show enhancements in current followed by deficits, whichever direction they are travelling away from the reconnection site. However, the leading peaks of the pressure perturbations show an enhancement when travelling across the width (Fig.~\ref{fig:vertcontsw}), but a deficit when travelling through the depth (Fig.~\ref{fig:vertcontsd}) of the current layer. The former are launched from the narrow edges of the current layer, as if from a point, and propagate outwards, in a spherical-like manner, into the cusp regions, which in the MHS equilibrium, have larger pressure than the regions outside the cusps. 

Outside the cusps, the latter type of perturbations are launched (i.e., carrying a deficit in pressure) from the comparatively wide edge of the current layer. These perturbations are more planar in nature, thus they move linearly outwards in a block from either side of the current layer. The left-hand columns of Figs.~\ref{fig:vertcontsd} and \ref{fig:vertcontsw} show that the perturbations in current and pressure show a consistent behaviour down the entire length of the separator current layer both through its depth and across its width.  Remember though that the current layer is twisted and so these perturbations emanate out along the separator forming an expanded helical pattern. This is shown in a movie, available online, where isosurfaces of the pressure difference (drawn at $p-p_0=1 \times 10^{-4}$ in yellow and $p-p_0=-1 \times 10^{-4}$ in blue) evolve through the reconnection experiment. In this movie, we are looking down on the 3D model and so only the negative null (red sphere) is seen but the spines of both nulls are visible. 

As these perturbed pulses travel further from the reconnection site they are followed by a second set of pulses, which show the same basic behaviour. These pulses are the ones that were launched at the same time as the lead pulses, but travelled inwards across the current layer, rather than outwards from it. Naturally, therefore, for the wave pulses travelling outwith the cusps, the following planar pulses are very close behind the lead planar ones since they simply cross the (thin) depth of the current layer (right-hand column of Figs.~\ref{fig:perpconts} and \ref{fig:vertcontsd}). In the cusps themselves, the following pulses that leave the narrow edges of the current layer have to cross the entire width of the current layer and, therefore, these are much further behind the leading point-like pulses (right-hand column of Figs.~\ref{fig:perpconts} and \ref{fig:vertcontsw}). 

\begin{figure}[h!]
\centerline{
\includegraphics[width=0.5\textwidth,clip]{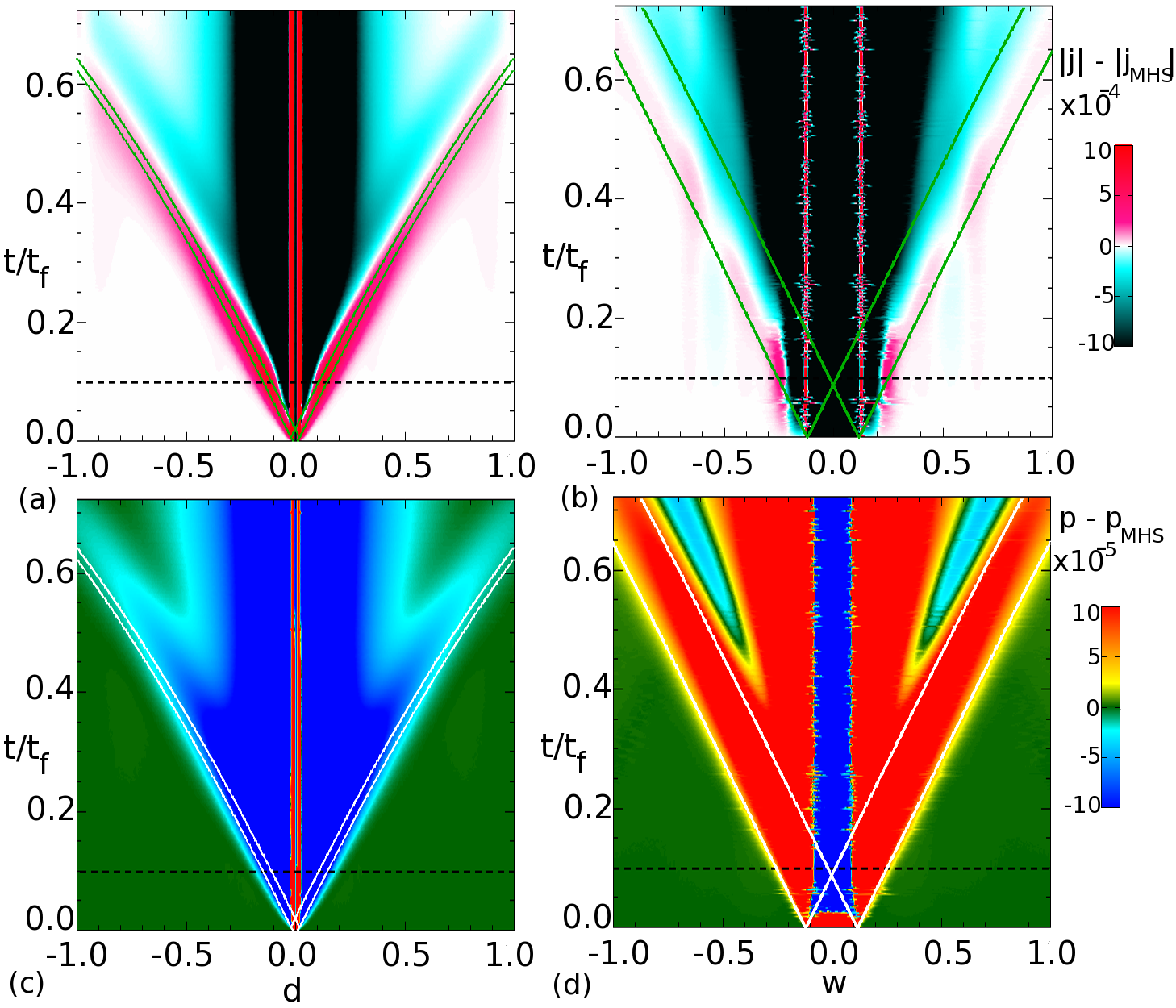}}
\caption{Time slices of the perturbed current (upper row) and the perturbed pressure (lower row) plotted through the depth (first column) and across the width (second column) of the current layer, in the plane at $z=0.4$. The black dashed lines highlight where the first phase ends and phase II begins. The green/white lines start on the edge of the current layer and represent a wave travelling at the fast-magnetoacoustic speed.}\label{fig:phase2waves}
\end{figure}

In Fig.~\ref{fig:phase2waves}, time slices of the perturbed current and pressure through the depth and across the width of the current layer in the $z=0.4$ plane are plotted. These show the wave pulses that travel from the edges of the diffusion region and match well with the over plotted lines that indicate the speed that fast-magnetoacoustic waves, launched both inwards and outwards from the edges of the diffusion region, would travel at. 

\subsection{Steady Flows}
In the wake of these perturbations, deficits of current exist both outside and within the cusps. Deficits of pressure exist only outwith the cusps, while enhancements of pressure inside the cusps form locally around the diffusion region (Figs. \ref{fig:perpconts}, \ref{fig:vertcontsd} and \ref{fig:vertcontsw}). These regions do not move at a particular wave speed, but expand out slowly throughout the duration of the experiment. Due to the loss of equilibrium within the separator current layer, the region surrounding the current layer, which is non-resistive, responds in order to try and regain force balance by trying to rebuild the current within the separator current layer. The forces present are basically the same as those found during the formation of the initial MHS equilibrium \citep{Stevenson15}. 

Outside the cusp regions, the inward directed magnetic pressure forces once more dominate over the outward directed plasma pressure force, generating an inflow towards the separator current layer in this region. Inside the cusp regions, outward directed magnetic tension forces dominate over the inward directed plasma pressure forces causing an outflow. These flows are maintained because, as soon as the current within the current layer strengthens to $|{\bf{j}}| = j_{crit}$, diffusivity dissipates this current and thus prevents a static equilibrium being formed. Instead, a system, which is close to a steady state, is created involving slow reconnection at the separator current layer (phase II of the reconnection process). 

\subsection{Infinite-Time Collapse}
Fig.~\ref{fig:perpconts}a shows that the current enhancements on the separatrix surfaces are also perturbed, but there is no corresponding perturbation in pressure at this time (Fig.~\ref{fig:perpconts}d). Furthermore, the current along the separatrix surfaces is not decreased, as one might expect if these perturbations were the result of reconnection, but instead is increased. 

Since the separatrix surfaces are outside the diffusion region, in an area where the magnetic plasma is non-resistive, the plasma continues to behave as it did in the relaxation experiment discussed in detail in \citet{Stevenson15}. In particular, we recall that our initial MHS equilibrium is not a true equilibrium, but is in force balance everywhere, except within the current layers on the separatrices and along the separator (the formation of a true equilibrium would take an infinite time). On the edges of the separatrix surfaces there are small residual forces that are very slowly increasing the current within the current layer and this is what continues to happen in our experiment here.

\subsection{Vorticity and Velocity}\label{sec:vortandvel}
In order to understand the nature of the flows created as a result of the reconnection, we consider both the vorticity and the velocity at three different times. The final row of graphs in Figs.~\ref{fig:perpconts}, \ref{fig:vertcontsd} and \ref{fig:vertcontsw} show the vorticity in the $z=0.4$ plane perpendicular to the separator, through the depth and across the width of the current layer, respectively. Over plotted on these graphs are arrows indicating the direction and size of the velocity in these cuts. The arrows are coloured depending on their position: the arrows are coloured black where $d \textless 0$ and $w \textless 0$ and the arrows are coloured green where $d \textgreater 0$ and $w \textgreater 0$. We have coloured the arrows this way so that the direction of the flow from a given side of the diffusion region is clearer. 

Figs.~\ref{fig:perpconts}g-\ref{fig:perpconts}i show a very similar pattern to the classical quadrupolar vortex scenario and stagnation flow found in 2D X-point reconnection regimes. The main difference is that instead of finding zero vorticity in the vicinity of the separator, an antiparallel flow is found associated with a clockwise (blue) rotating flow pattern (Fig.~\ref{fig:perpconts}i). 

If instead of considering the $z=0.4$ plane, we looked at the vortex pattern in the $z=0.6$ plane (see movie of Fig.~\ref{fig:perpconts}), then we would find a similar quadrupole vortex (rotated slightly due to the twisted nature of the current layer and separartix surfaces about the separator), but in the vicinity of the separator, an antiparallel flow associated with an anticlockwise (red) rotation would be found. This agrees with the existence of a counter-rotating flow along the separator discussed in \citet{Stevenson15_jgra}. 

Not surprisingly, therefore, looking at the vorticity in the cut through the depth (Fig.~\ref{fig:vertcontsd}) and across the width (Fig.~\ref{fig:vertcontsw}) of the current layer, we see that the directions of the flows change with position along the separator. The velocity arrows indicate that the dominant flows are directed inwards through the depth, and outwards across the width, for all $z$. However, superimposed on these are weak stagnation-type flow patterns. 

Along the separator, in the cut through the depth (Figs.~\ref{fig:vertcontsd}g-\ref{fig:vertcontsd}i), weak flows run towards both nulls from a point 0.6 times the length of the separator (as measured from the lower null). The location of this “stagnation point”, which corresponds to where the flows are purely directed inwards to the current layer through the depth, does not appear to move over time. Its location is a result of the fact that the plasma pressure on the separator is greatest at this point in the MHS equilibrium and so the strongest magnetic pressure force must have existed at this location to counter the largest pressure force that would have been located there.

The cuts across the width of the current layer (Figs.~\ref{fig:vertcontsw}g-\ref{fig:vertcontsw}i) show the opposite quadrupolar-vortex pattern close to the separator. This pattern shows that there are weak flows that run in from the nulls along the separator to a point 0.4 times the length of the separator (as measured from the lower null) at time $t=0.019t_f$ (Fig.~\ref{fig:vertcontsw}g). This is the approximate centre of the main stagnation outflow across the width of the current layer. It also corresponds to the point where the current is largest along the separator in the MHS equilibrium, and thus where the outwardly directed tension force must have been highest. So, as soon as the equilibrium is lost, a strong outflow in the $z=0.4$ plane results. 

Unlike the stagnation flow through the depth of the current layer, the stagnation-point flow across the width changes over time, such that at the time phase I ends, there seem to be two stagnation points on the separator. During phase II a single stagnation-type flow has reformed. It is not clear what the multiple stagnation-point flows indicate, but it is important to remember that at the transition between the two phases very little reconnection, if any, is occurring since the slow, impulsive-bursty reconnection phase has not yet formed. 

\section{Transport of Energy}\label{sec:transp}
We have already seen that the reconnection at the separator current layer leads to waves being launched in the system, from the edges of the diffusion site, due to the sudden lack of force balance. These waves travel out and cause the magnetic field and plasma to change, setting up flows in the system. In this section, we analyse the transport of magnetic, internal and kinetic energy equations (as detailed in \citet{Birn09}), integrated over volumes within our domain, in order to see what quantities these waves and flows carry with them.

\begin{figure}[h!]
\centerline{\includegraphics[width=0.25\textwidth,clip]{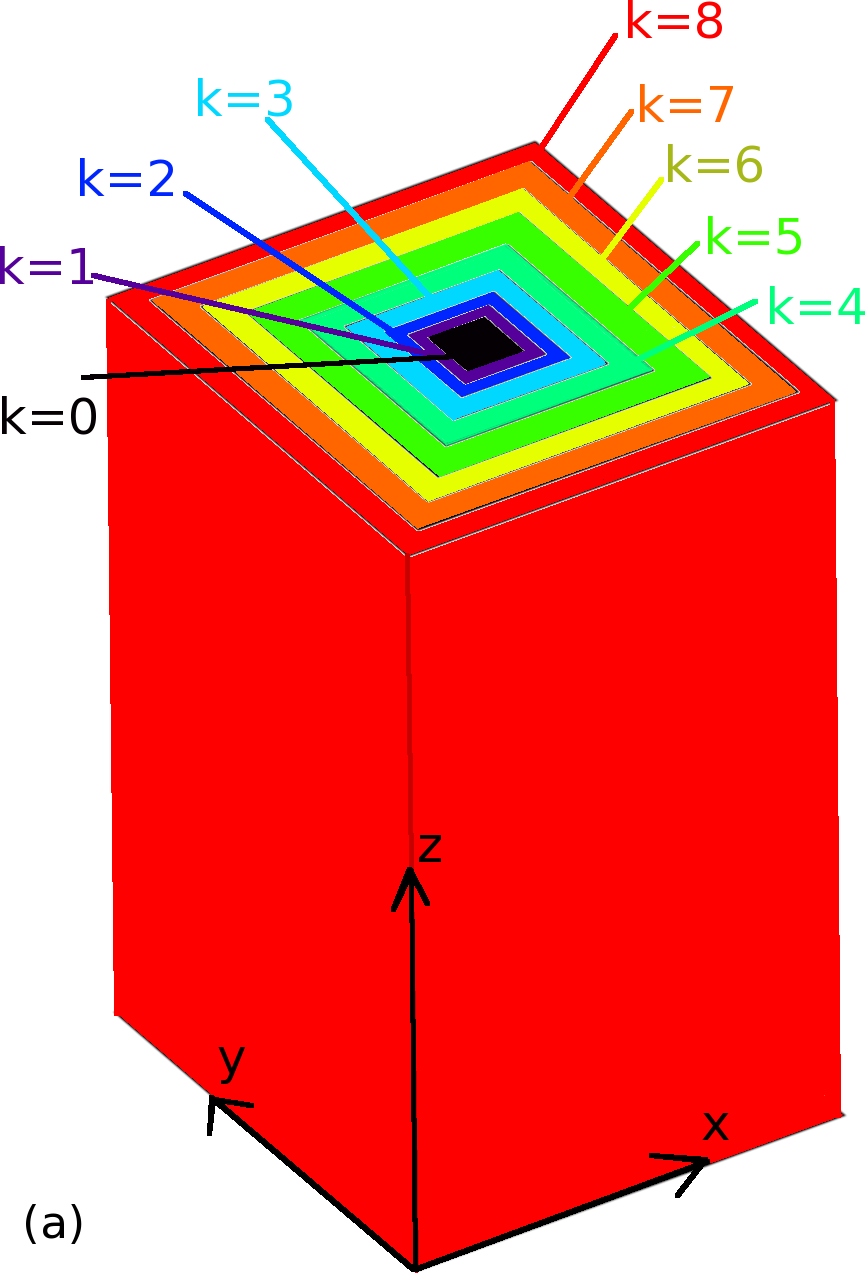}
\includegraphics[width=0.25\textwidth,clip]{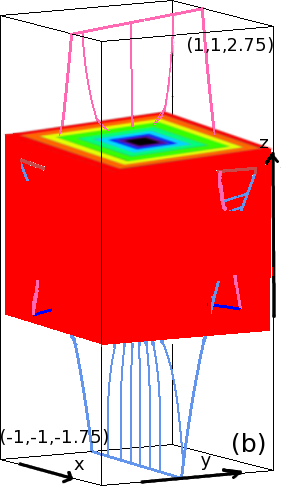}}
\caption{(a) Cartoon depicting the nine volumes over which the transport of energy equations are integrated. The volumes increase according to $-(0.15+k/10) \le x,y \le (0.15+k/10)$ and $-0.2 \le z \le 1.2$ shown by the colours black ($k=0$), purple ($k=1$), blue ($k=2$), cyan ($k=3$), lime ($k=4$), green ($k=5$), yellow ($k=6$), orange ($k=7$) and red ($k=8$), with the last box being just smaller than the size of the domain in $x$ and $y$. (b) For context, the MHS equilibrium skeleton with boxes overdrawn.}\label{fig:transportbox}
\end{figure}

To determine the quantities in the transport equations, we integrate each of them separately over sub-volumes that increase in size within our domain. The nine volumes have sizes $-(0.15+k/10) \le x,y \le (0.15+k/10)$, for $k=0,1,2,...,8$ and the $z$-range is fixed for all volumes at $-0.2 \le z \le 1.2$, since the waves and flows travel horizontally and not vertically out from the separator. Therefore, the smallest volume encloses the current layer, the second volume encloses the first volume and so on up to the largest volume, which is slightly smaller than the domain size in the $x$ and $y$ dimensions. A cartoon of these volumes is shown in Fig.~\ref{fig:transportbox}a, where the volumes are coloured black, purple, blue, lime, green, yellow, orange and red as they increase in size. These volumes are shown drawn with the MHS equilibrium skeleton in Fig.~\ref{fig:transportbox}b to highlight the size of the boxes compared to the skeleton. Hence, in each plot of Fig.~\ref{fig:transportbik}, which shows the time evolution of the energy transport quantities, there are nine lines coloured to match these volumes. 

\subsection{Transport of Magnetic Energy}
The transport of magnetic energy equation states that
\begin{equation}
\frac{\partial}{\partial t}\Bigg(\frac{B^2}{2\mu_0}\Bigg) = -\eta j^2-\nabla \cdot({\bf{E}} \times {\bf{B}}) -{\bf{v}}\cdot({\bf{j}} \times {\bf{B}}),\label{eq:transb}
\end{equation}
where $t$ is time, $B^2$ is the square of the magnitude of the magnetic field ($B = |{\bf{B}}|$), $\mu_0$ is the magnetic permeability which is equal to one in our dimensionless units, $j^2$ is the square of the current magnitude ($j=|{\bf{j}}|$) and ${\bf{E}}$ is the electric field.
Hence, the rate of change of magnetic energy, throughout the reconnection experiment (Fig.~\ref{fig:transportbik}a), is made up of the negative sum of the Ohmic dissipation (Fig.~\ref{fig:transportbik}b), the Poynting flux (Fig.~\ref{fig:transportbik}c) and the work done by the Lorentz force (Fig.~\ref{fig:transportbik}d). 

At the start of the experiment, there is an immediate drop in the rate of change of magnetic energy corresponding to the strong Ohmic dissipation during phase I, the fast-reconnection phase (Fig.~\ref{fig:transportbik}b). The integrated values of the Ohmic dissipation, which are of the order $10^{-3}$, are the same regardless of the size of the volume (so only one line is visible) since the Ohmic dissipation occurs in the diffusion region, which is enclosed in all boxes. The Ohmic dissipation is fairly constant after $t=0.09t_f$; the slow, impulsive-bursty second phase.

The value of the Poynting flux, integrated over all volumes, is initially positive indicating that the waves are travelling out through the boundaries of our volumes (Fig.~\ref{fig:transportbik}c). The waves then cause the plasma to change, setting up flows in the system near to the original diffusion site. The flows bring Poynting flux in through the smaller volumes, from about $t=0.16t_f$ onwards. However, the Poynting flux, carried out through the volumes by the waves, becomes relatively large the further out they travel (see the green to red lines in Fig.~\ref{fig:transportbik}c). Note, however, that the amount of Poynting flux is roughly 25 times smaller than the peak Ohmic dissipation in phase I.

The Lorentz force is working to try to regain force balance in the system from the moment the current in the separator current layer begins to be dissipated (positive values in Fig.~\ref{fig:transportbik}d). This figure shows that the work done by the Lorentz force, which is roughly of the order of $3 \times 10^{-5}$, is acting out through the sub-volumes over which we have integrated. This is directly related to the direction of the magnetic tension and magnetic pressure forces which make up the Lorentz force. The magnetic tension force, which acts to straighten the field lines, is directed outwards from the diffusion site both within and outwith the cusp regions which are formed by the separatrix surfaces of the nulls. The magnetic pressure force is directed in towards the diffusion region within the cusps and outwith the cusps. Overall, these forces sum such that the work done by the Lorentz force is acting outwards away from the diffusion region. The magnitude of this term is roughly 30 times smaller than the Ohmic dissipation term.

\subsection{Transport of Internal Energy}
Eq.~\ref{eq:transi} is the transport of internal energy equation
\begin{equation}
\frac{\partial}{\partial t}(\rho \epsilon) = \eta j^2 - \nabla \cdot \Big((p + \rho \epsilon){\bf{v}}\Big) - (-{\bf{v}}\cdot \nabla p),\label{eq:transi}
\end{equation}
where $\rho$ is the density and $\epsilon$ is the internal energy per unit mass.

\begin{figure*}[h!]
\centerline{
\includegraphics[width=0.33\textwidth,clip]{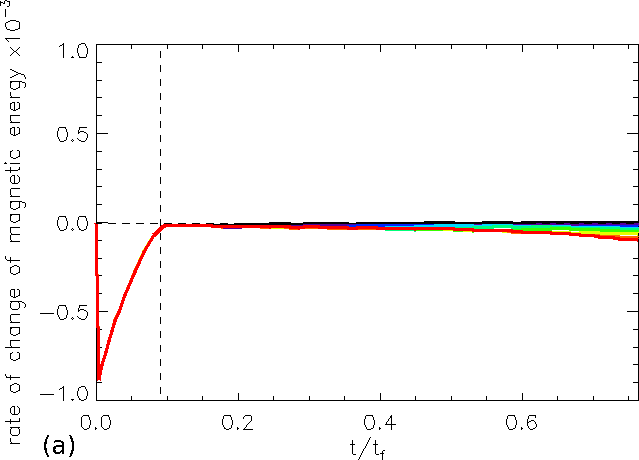}
\includegraphics[width=0.33\textwidth,clip]{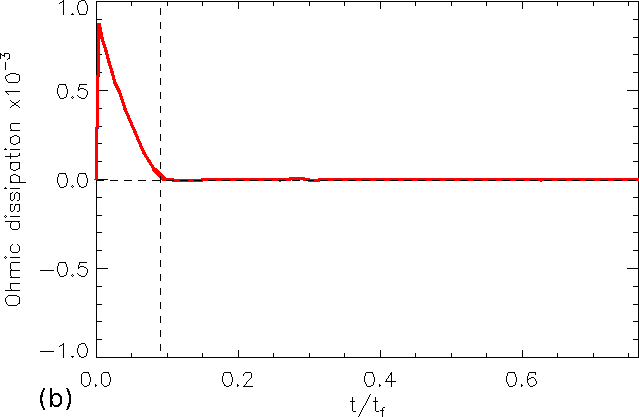}
\includegraphics[width=0.33\textwidth,clip]{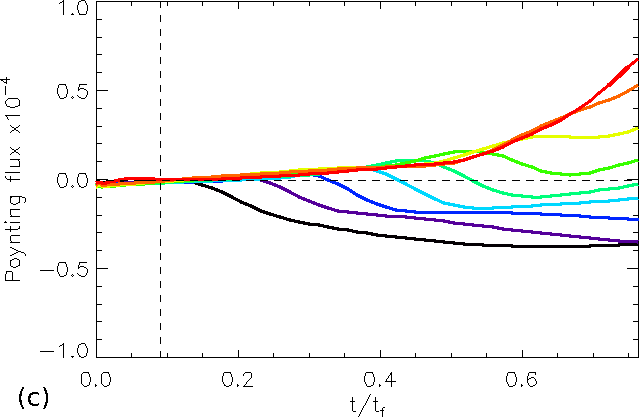}}
\centerline{
\includegraphics[width=0.33\textwidth,clip]{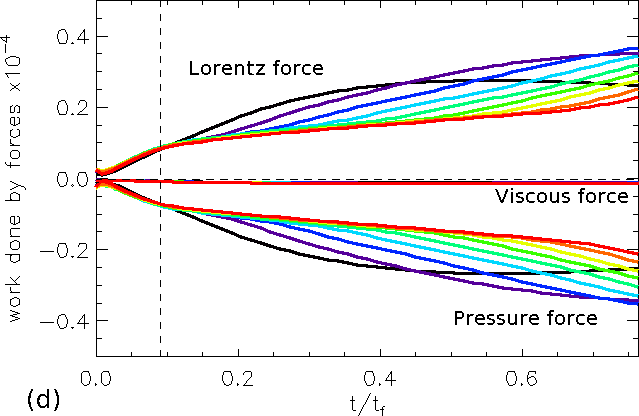}
\includegraphics[width=0.33\textwidth,clip]{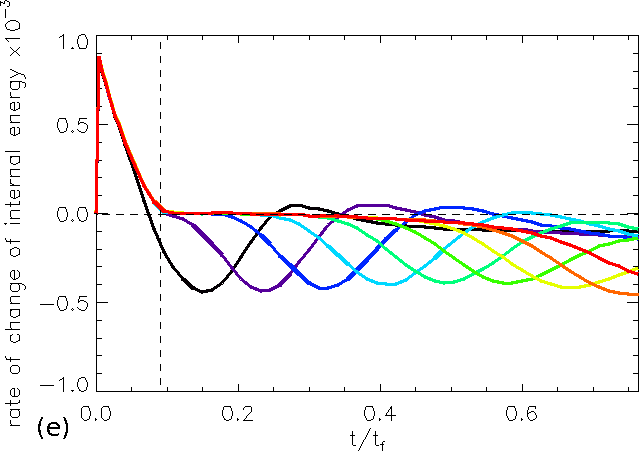}
\includegraphics[width=0.33\textwidth,clip]{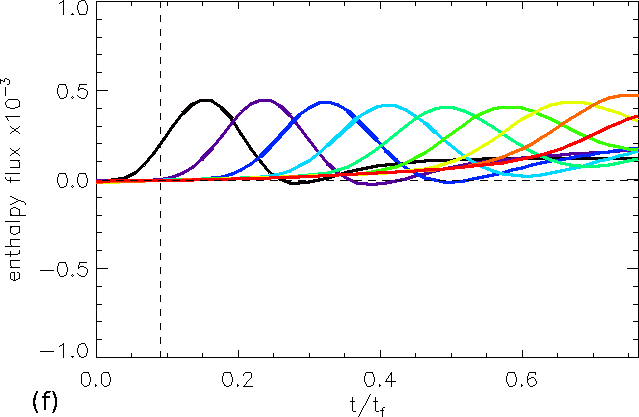}}
\caption{Quantities, plotted against time, of (a) the rate of change of the magnetic energy, (b) the Ohmic dissipation, (c) the Poynting flux, (d) the work done by the Lorentz, viscous and pressure forces, (e) the rate of change of internal energy and (f) the enthalpy flux. The line colour represents the different volumes over which these quantities have been integrated (c.f. Fig~\ref{fig:transportbox}). The black dashed vertical line highlights where phase I ends and phase II begins and the black dashed horizontal line indicates zero.}\label{fig:transportbik}
\end{figure*}

Here, we can write $\rho \epsilon=3p/2$, since our closure equation is $\epsilon=p/\rho(\gamma-1)$, and $\gamma=5/3$ and therefore $p+ \rho \epsilon=5p/2$. This equation states that the rate of change of internal energy is due to the Ohmic heating minus the enthalpy flux minus the work done by the pressure force. Fig.~\ref{fig:transportbik}e shows the rate of change of internal energy, which is of the order of $10^{-3}$, integrated over all nine volumes, throughout the experiment. The initial sharp spike in this figure is due to the Ohmic heating (Fig.~\ref{fig:transportbik}b) as was seen in the energetics discussed in \citet{Stevenson15_jgra}. 

After this spike, the rate of change of internal energy decreases and becomes negative. A small travelling wave is seen moving out through all the sub-volumes which comes from the enthalpy flux term (Fig.~\ref{fig:transportbik}f). This term is of the order of $5 \times 10^{-4}$ and is half the size of the peak Ohmic dissipation during phase I. 

The final term which contributes to Eq.~\ref{eq:transi} is the work done by the pressure force (negative values in Fig.~\ref{fig:transportbik}d), however, the magnitude of this term is about 30 times smaller than that of the Ohmic heating and about 15 times smaller than the enthalpy flux and so its contribution is small here. The work done by the pressure force is directed in through the sub-volumes over which we integrate and is acting, like the Lorentz force, to try to regain force balance in the system as soon as the reconnection begins.

\subsection{Transport of Kinetic Energy}
The transport of kinetic energy equation states that the rate of change of kinetic energy is equal to the work done by the Lorentz force plus the work done by the pressure force plus the work done by the viscous force minus the bulk kinetic energy flux
\begin{equation}
\frac{\partial}{\partial t}\Big(\frac{\rho v^2}{2}\Big) = {\bf{v}}\cdot({\bf{j}} \times {\bf{B}}) + (-{\bf{v}}\cdot \nabla p) + {\bf{v}}\cdot{\bf{F_{\nu}}}-\nabla \cdot \Big(\frac{\rho v^2}{2}{\bf{v}}\Big),
\end{equation}
where ${\textbf{F}}_{\nu} = \nu(\nabla^2{\textbf{v}} + \tfrac{1}{3}\nabla(\nabla \cdot {\textbf{v}}))$ is the viscous force.

The rate of change of kinetic energy is very small ($\sim5 \times 10^{-7}$) throughout the reconnection experiment. This is because the work done by the Lorentz and pressure forces (Figs.~\ref{fig:transportbik}d) are about equal in size, but are of opposite sign (they are acting to regain force balance in the system after the dissipation of the current layer). Also, contributions from the work done by the viscous force (values close to zero in Fig.~\ref{fig:transportbik}d) and the bulk kinetic energy flux are very small ($\sim5 \times 10^{-7}$) since the velocities in the system have small magnitudes.

Overall, we have found that there are five main terms which play a significant role in the transport of energy in our experiment. Ohmic heating plays the most significant role in our experiment, especially during phase I, converting magnetic energy into internal energy. This energy is then carried away from the diffusion region by the enthalpy flux, which is half the size of the Ohmic heating term and the Poynting flux, which is roughly 25 times smaller than the peak Ohmic heating. The final two important terms are the work done by the Lorentz and pressure forces, which are similar in magnitude, but act in opposite directions and are both about 30 times smaller than the Ohmic heating term.

\section{Conclusions}\label{sec:conclusions}
In this paper, we have studied the properties of the waves and flows created due to spontaneous reconnection at a 3D separator current layer. We start with a system that is in MHS equilibrium everywhere save for very small forces at the current enhancements about the separator and separatrix surfaces. An anomalous diffusivity is applied such that reconnection only occurs at the separator current layer that twists about the separator. 

The onset of the reconnection produces waves that propagate out from the edge of the diffusion site at the separator current layer. These waves only have small amplitudes, due to the relatively small reconnection event that initiates them, and they travel at the fast-magnetoacoustic speed (which in our high-beta experiment is approximately equal to the slow-magnetoacoustic speed). 

They carry the dissipated current away from the diffusion region and disperse as they travel. The nature of the waves has the same pattern in all planes perpendicular to the separator, which is basically the same as that found due to waves launched following reconnection at a 2D X-type null. 

(i) Planar-like waves that are twisted about the separator are launched from either side of the diffusion region and travel away from the separator current layer carrying current and causing a deficit in pressure. Equivalent waves are also launched inwards through the depth of the current layer at the same time. These waves end up running closely behind the outwardly-launched waves.

(ii) Point-like waves that again are twisted about the separator are launched outwards from the narrow edges of the diffusion region. In any given plane perpendicular to the separator, these spread in a circular pattern carrying current away from the separator current layer and causing an enhancement in pressure.  As above, point-like waves also travel inwards across the width of the separator current layer, which is roughly 20 times the size of the depth, thus these waves lag behind the outward waves.

These waves communicate the sudden loss of force balance within the separator current layer and, hence, in their wake magnetic and plasma forces are set up with the aim of restoring the equilibrium. As already explained by \citet{Stevenson15}, an equilibrium in such a system with a separator involves a current enhancement about the separator and, thus, a velocity flow pattern is created, which brings in more flux from outwith the cusp regions to enhance the current at the separator. As soon as the current in the layer reaches the level of $j_{crit}$ the anomalous diffusivity dissipates it. This leads to a slow, impulsive-bursty second phase of reconnection driven by this stagnation-like flow. Although the reconnection during this phase is very slow, the Ohmic dissipation associated with it is still larger than the viscous heating associated with the damping of the magnetoacoustic waves and flows.

The amplitude of the waves that result from the reconnection are small and of the order $10^{-4}$. They are, however, much bigger (100 times) than those found in the 2D high-beta experiment of \citet{FFP12} which we believe is due to the third dimension permitting a larger current layer to be formed. To form an even larger current layer in our high-beta scenario we could have start, prior to forming the MHS equilibrium, with an initial magnetic field that has either greater initial current or different magnetic field parameters (see \citet{Stevenson15PhD} for full details). However, in the resulting MHS equilibrium the current layer was not resolved and we were concerned that numerical diffusion had occurred, in some cases, before the MHS equilibrium could be formed. Also, the resulting waves were only marginally greater in amplitude. Lowering the value of $j_{crit}$, the level above which the diffusivity is non-zero, would define a larger current layer, but this has the side effect of permitting reconnection within the current enhancements along the separatrix surfaces leading to more (complicated) wave pulses.

An analysis of the energy transport in the model shows that the Ohmic dissipation is about twice that of the enthalpy flux carried by the magnetoacoustic waves away from the reconnection site. In turn, the enthalpy flux is more than ten times the work done by either the Lorentz or the pressure forces (which are basically the same size, but cancel each other out since they work in opposite directions). The Poynting flux is also about ten times smaller than the enthalpy flux. The dominance of the enthalpy flux over the Poynting flux is not surprising since our experiments are high beta (as shown in \citet{Birn09}).

In order to compare our dimensionless results to those of a specific space plasma scenario, the speed of the waves generated by the reconnection can be scaled by the factor $B_{n}/\sqrt{\mu \rho_{n}}$. In situ measurements by Double Star have detected number densities of $n_n = 10 \times 10^6$ m$^{-3}$ and magnetic field strengths of $B_n=30$ nT at the magnetopause \citep{Trenchi2008}. Using a mean particle mass (this value has been calculated using magnetosheath abundances \citep{Gloeckler87} of the major magnetospheric ions \citep{Reme01}) of $m =1.07\times m_p$, where $m_p$ is the mass of a proton, this number density corresponds to a mass density of $\rho_n=1.8 \times 10^{-20}$ kgm$^{-3}$. In our experiment, the maximum dimensionless Alfv\'en speed in the outflow regions is $v_A=1.6$, thus we find that the maximum Alfv\'en speed in our domain is $v_A=320$ kms$^{-1}$. This is of the same order as the hybrid Alfv\'en speed, $v_{Ah}=380$ kms$^{-1}$, found between the magnetosheath and the magnetosphere in \citet{Komar13}. Note that, in our model, the density in the outflow region decreases away from the separator and the magnetic field strength increases, so if our domain was larger we would be able to measure greater values of the Alfv\'en speed than we have detailed here.

All the experiments discussed here can only be run until the travelling waves near the boundaries of the box. If it was possible to run these experiments for longer then the viscous heating may increase sufficiently to become comparable with the observed Ohmic heating in this second phase. Furthermore, a low-beta system may also permit greater viscous heating. We plan, in a follow up paper, to investigate if this is the case. 

\begin{acknowledgments}
JEHS would like to thank STFC for financial support during her Ph.D and continued support after on the St Andrews SMTG's STFC consortium grant. CEP also acknowledges support from this same grant. Computations were carried out on the UKMHD consortium cluster funded by STFC and SRIF. Data from simulation results are available on request from J. E. H. Stevenson (email: jm686@st-andrews.ac.uk).
\end{acknowledgments}

\providecommand{\noopsort}[1]{}\providecommand{\singleletter}[1]{#1}%


\providecommand{\noopsort}[1]{}\providecommand{\singleletter}[1]{#1}%
\begin{thebibliography}{68}
\providecommand{\natexlab}[1]{#1}
\expandafter\ifx\csname urlstyle\endcsname\relax
  \providecommand{\doi}[1]{doi:\discretionary{}{}{}#1}\else
  \providecommand{\doi}{doi:\discretionary{}{}{}\begingroup
  \urlstyle{rm}\Url}\fi

\bibitem[{\textit{{Alexeev} et~al.}(1998)\textit{{Alexeev}, {Sibeck}, and
  {Bobrovnikov}}}]{Alexeev98}
{Alexeev}, I.~I., D.~G. {Sibeck}, and S.~Y. {Bobrovnikov} (1998), {Concerning
  the location of magnetopause merging as a function of the magnetopause
  current strength}, \textit{\jgr}, \textit{103}, 6675--6684,
  \doi{10.1029/97JA02863}.

\bibitem[{\textit{{Arber} et~al.}(2001)\textit{{Arber}, {Longbottom},
  {Gerrard}, and {Milne}}}]{Arber01}
{Arber}, T.~D., A.~W. {Longbottom}, C.~L. {Gerrard}, and A.~M. {Milne} (2001),
  {A Staggered Grid, Lagrangian-Eulerian Remap Code for 3-D MHD Simulations},
  \textit{Journal of Computational Physics}, \textit{171}, 151--181,
  \doi{10.1006/jcph.2001.6780}.

\bibitem[{\textit{{Arzner} and {Scholer}}(2001)}]{Arzner01}
{Arzner}, K., and M.~{Scholer} (2001), {Kinetic structure of the post plasmoid
  plasma sheet during magnetotail reconnection}, \textit{\jgr}, \textit{106},
  3827--3844, \doi{10.1029/2000JA000179}.

\bibitem[{\textit{{Birn} et~al.}(2001)\textit{{Birn}, {Drake}, {Shay},
  {Rogers}, {Denton}, {Hesse}, {Kuznetsova}, {Ma}, {Bhattacharjee}, {Otto}, and
  {Pritchett}}}]{Birn01}
{Birn}, J., J.~F. {Drake}, M.~A. {Shay}, B.~N. {Rogers}, R.~E. {Denton},
  M.~{Hesse}, M.~{Kuznetsova}, Z.~W. {Ma}, A.~{Bhattacharjee}, A.~{Otto}, and
  P.~L. {Pritchett} (2001), {Geospace Environmental Modeling (GEM) magnetic
  reconnection challenge}, \textit{\jgr}, \textit{106}, 3715--3720,
  \doi{10.1029/1999JA900449}.

\bibitem[{\textit{{Birn} et~al.}(2009)\textit{{Birn}, {Fletcher}, {Hesse}, and
  {Neukirch}}}]{Birn09}
{Birn}, J., L.~{Fletcher}, M.~{Hesse}, and T.~{Neukirch} (2009), {Energy
  Release and Transfer in Solar Flares: Simulations of Three-Dimensional
  Reconnection}, \textit{\apj}, \textit{695}, 1151--1162,
  \doi{10.1088/0004-637X/695/2/1151}.

\bibitem[{\textit{{Biskamp}}(2000)}]{Biskamp2000}
{Biskamp}, D. (2000), \textit{{Magnetic Reconnection in Plasmas}}.

\bibitem[{\textit{{Borovsky}}(2008)}]{Borovsky08}
{Borovsky}, J.~E. (2008), {The rudiments of a theory of solar
  wind/magnetosphere coupling derived from first principles}, \textit{Journal
  of Geophysical Research (Space Physics)}, \textit{113}, A08228,
  \doi{10.1029/2007JA012646}.

\bibitem[{\textit{{Borovsky}}(2013)}]{Borovsky13}
{Borovsky}, J.~E. (2013), {Physical improvements to the solar wind reconnection
  control function for the Earth's magnetosphere}, \textit{Journal of
  Geophysical Research (Space Physics)}, \textit{118}, 2113--2121,
  \doi{10.1002/jgra.50110}.

\bibitem[{\textit{{Cao} et~al.}(2013)\textit{{Cao}, {Wei}, {Duan}, {Fu},
  {Zhang}, {Reme}, and {Dandouras}}}]{Cao13}
{Cao}, J.~B., X.~H. {Wei}, A.~Y. {Duan}, H.~S. {Fu}, T.~L. {Zhang}, H.~{Reme},
  and I.~{Dandouras} (2013), {Slow magnetosonic waves detected in reconnection
  diffusion region in the Earth's magnetotail}, \textit{Journal of Geophysical
  Research (Space Physics)}, \textit{118}, 1659--1666,
  \doi{10.1002/jgra.50246}.

\bibitem[{\textit{{Close} et~al.}(2005)\textit{{Close}, {Parnell}, {Longcope},
  and {Priest}}}]{Close05}
{Close}, R.~M., C.~E. {Parnell}, D.~W. {Longcope}, and E.~R. {Priest} (2005),
  {Coronal Flux Recycling Times}, \textit{Solar Physics}, \textit{231}, 45--70,
  \doi{10.1007/s11207-005-6878-1}.

\bibitem[{\textit{{Dorelli} and {Bhattacharjee}}(2008)}]{Dorelli08}
{Dorelli}, J.~C., and A.~{Bhattacharjee} (2008), {Defining and identifying
  three-dimensional magnetic reconnection in resistive magnetohydrodynamic
  simulations of Earth's magnetospherea)}, \textit{Physics of Plasmas},
  \textit{15}(5), 056504, \doi{10.1063/1.2913548}.

\bibitem[{\textit{{Dorelli} and {Bhattacharjee}}(2009)}]{Dorelli09}
{Dorelli}, J.~C., and A.~{Bhattacharjee} (2009), {On the generation and
  topology of flux transfer events}, \textit{Journal of Geophysical Research
  (Space Physics)}, \textit{114}, A06213, \doi{10.1029/2008JA013410}.

\bibitem[{\textit{{Drake} et~al.}(1997)\textit{{Drake}, {Biskamp}, and
  {Zeiler}}}]{Drake97}
{Drake}, J.~F., D.~{Biskamp}, and A.~{Zeiler} (1997), {Breakup of the electron
  current layer during 3-D collisionless magnetic reconnection}, \textit{\grl},
  \textit{24}, 2921--2924, \doi{10.1029/97GL52961}.

\bibitem[{\textit{{Fuentes-Fern{\'a}ndez}
  et~al.}(2011)\textit{{Fuentes-Fern{\'a}ndez}, {Parnell}, and {Hood}}}]{FF11}
{Fuentes-Fern{\'a}ndez}, J., C.~E. {Parnell}, and A.~W. {Hood} (2011),
  {Magnetohydrodynamics dynamical relaxation of coronal magnetic fields. II. 2D
  magnetic X-points}, \textit{Astronomy and Astrophysics}, \textit{536}, A32,
  \doi{10.1051/0004-6361/201117156}.

\bibitem[{\textit{{Fuentes-Fern{\'a}ndez}
  et~al.}(2012{\natexlab{a}})\textit{{Fuentes-Fern{\'a}ndez}, {Parnell},
  {Hood}, {Priest}, and {Longcope}}}]{FFP12}
{Fuentes-Fern{\'a}ndez}, J., C.~E. {Parnell}, A.~W. {Hood}, E.~R. {Priest}, and
  D.~W. {Longcope} (2012{\natexlab{a}}), {Consequences of spontaneous
  reconnection at a two-dimensional non-force-free current layer},
  \textit{Physics of Plasmas}, \textit{19}(2), 022,901,
  \doi{10.1063/1.3683002}.

\bibitem[{\textit{{Fuentes-Fern{\'a}ndez}
  et~al.}(2012{\natexlab{b}})\textit{{Fuentes-Fern{\'a}ndez}, {Parnell}, and
  {Priest}}}]{FFP12b}
{Fuentes-Fern{\'a}ndez}, J., C.~E. {Parnell}, and E.~R. {Priest}
  (2012{\natexlab{b}}), {The onset of impulsive bursty reconnection at a
  two-dimensional current layer}, \textit{Physics of Plasmas}, \textit{19}(7),
  072,901, \doi{10.1063/1.4729334}.

\bibitem[{\textit{{Fujimoto} and {Sydora}}(2008)}]{Fujimoto08}
{Fujimoto}, K., and R.~D. {Sydora} (2008), {Whistler waves associated with
  magnetic reconnection}, \textit{\grl}, \textit{35}, L19112,
  \doi{10.1029/2008GL035201}.

\bibitem[{\textit{{Galsgaard} and {Nordlund}}(1997)}]{GN97}
{Galsgaard}, K., and {\AA}.~{Nordlund} (1997), {Heating and activity of the
  solar corona. 3. Dynamics of a low beta plasma with three-dimensional null
  points}, \textit{\jgr}, \textit{102}, 231--248, \doi{10.1029/96JA02680}.

\bibitem[{\textit{{Galsgaard} et~al.}(2000)\textit{{Galsgaard}, {Priest}, and
  {Nordlund}}}]{G00}
{Galsgaard}, K., E.~R. {Priest}, and {\AA}.~{Nordlund} (2000),
  {Three-dimensional Separator Reconnection - How Does It Occur?},
  \textit{Solar Physics}, \textit{193}, 1--16, \doi{10.1023/A:1005248811680}.

\bibitem[{\textit{{Gloeckler} and {Hamilton}}(1987)}]{Gloeckler87}
{Gloeckler}, G., and D.~C. {Hamilton} (1987), {AMPTE ion composition results},
  \textit{Physica Scripta Volume T}, \textit{18}, 73--84,
  \doi{10.1088/0031-8949/1987/T18/009}.

\bibitem[{\textit{{Gonzalez} and {Mozer}}(1974)}]{Gonzalez74}
{Gonzalez}, W.~D., and F.~S. {Mozer} (1974), {A quantitative model for the
  potential resulting from reconnection with an arbitrary interplanetary
  magnetic field}, \textit{\jgr}, \textit{79}, 4186--4194,
  \doi{10.1029/JA079i028p04186}.

\bibitem[{\textit{{Gosling} et~al.}(1986)\textit{{Gosling}, {Thomsen}, {Bame},
  and {Russell}}}]{Gosling86}
{Gosling}, J.~T., M.~F. {Thomsen}, S.~J. {Bame}, and C.~T. {Russell} (1986),
  {Accelerated plasma flows at the near-tail magnetopause}, \textit{\jgr},
  \textit{91}, 3029--3041, \doi{10.1029/JA091iA03p03029}.

\bibitem[{\textit{{Hara} et~al.}(2011)\textit{{Hara}, {Watanabe}, {Harra},
  {Culhane}, and {Young}}}]{Hara11}
{Hara}, H., T.~{Watanabe}, L.~K. {Harra}, J.~L. {Culhane}, and P.~R. {Young}
  (2011), {Plasma Motions and Heating by Magnetic Reconnection in a 2007 May 19
  Flare}, \textit{\apj}, \textit{741}, 107, \doi{10.1088/0004-637X/741/2/107}.

\bibitem[{\textit{{Haynes} et~al.}(2007)\textit{{Haynes}, {Parnell},
  {Galsgaard}, and {Priest}}}]{Haynes07}
{Haynes}, A.~L., C.~E. {Parnell}, K.~{Galsgaard}, and E.~R. {Priest} (2007),
  {Magnetohydrodynamic evolution of magnetic skeletons}, \textit{Royal Society
  of London Proceedings Series A}, \textit{463}, 1097--1115,
  \doi{10.1098/rspa.2007.1815}.

\bibitem[{\textit{{Hesse} and {Schindler}}(1988)}]{Hesse88}
{Hesse}, M., and K.~{Schindler} (1988), {A theoretical foundation of general
  magnetic reconnection}, \textit{Journal of Geophysical Research},
  \textit{93}, 5559--5567, \doi{10.1029/JA093iA06p05559}.

\bibitem[{\textit{{Hesse} et~al.}(2013)\textit{{Hesse}, {Aunai}, {Zenitani},
  {Kuznetsova}, and {Birn}}}]{Hesse13}
{Hesse}, M., N.~{Aunai}, S.~{Zenitani}, M.~{Kuznetsova}, and J.~{Birn} (2013),
  {Aspects of collisionless magnetic reconnection in asymmetric systems},
  \textit{Physics of Plasmas}, \textit{20}(6), 061210, \doi{10.1063/1.4811467}.

\bibitem[{\textit{{Hoshino} et~al.}(1998)\textit{{Hoshino}, {Mukai},
  {Yamamoto}, and {Kokubun}}}]{Hoshino98}
{Hoshino}, M., T.~{Mukai}, T.~{Yamamoto}, and S.~{Kokubun} (1998), {Ion
  dynamics in magnetic reconnection: Comparison between numerical simulation
  and Geotail observations}, \textit{\jgr}, \textit{103}, 4509--4530,
  \doi{10.1029/97JA01785}.

\bibitem[{\textit{{Innes} et~al.}(1997)\textit{{Innes}, {Inhester}, {Axford},
  and {Wilhelm}}}]{Innes97}
{Innes}, D.~E., B.~{Inhester}, W.~I. {Axford}, and K.~{Wilhelm} (1997),
  {Bi-directional plasma jets produced by magnetic reconnection on the Sun},
  \textit{Nature}, \textit{386}, 811--813, \doi{10.1038/386811a0}.

\bibitem[{\textit{{Ko} et~al.}(2003)\textit{{Ko}, {Raymond}, {Lin}, {Lawrence},
  {Li}, and {Fludra}}}]{Ko03}
{Ko}, Y.-K., J.~C. {Raymond}, J.~{Lin}, G.~{Lawrence}, J.~{Li}, and A.~{Fludra}
  (2003), {Dynamical and Physical Properties of a Post-Coronal Mass Ejection
  Current Sheet}, \textit{\apj}, \textit{594}, 1068--1084,
  \doi{10.1086/376982}.

\bibitem[{\textit{{Komar} et~al.}(2013)\textit{{Komar}, {Cassak}, {Dorelli},
  {Glocer}, and {Kuznetsova}}}]{Komar13}
{Komar}, C.~M., P.~A. {Cassak}, J.~C. {Dorelli}, A.~{Glocer}, and M.~M.
  {Kuznetsova} (2013), {Tracing magnetic separators and their dependence on IMF
  clock angle in global magnetospheric simulations}, \textit{Journal of
  Geophysical Research (Space Physics)}, \textit{118}, 4998--5007,
  \doi{10.1002/jgra.50479}.

\bibitem[{\textit{{Lau} and {Finn}}(1990)}]{LauFinn}
{Lau}, Y.-T., and J.~M. {Finn} (1990), {Three-dimensional kinematic
  reconnection in the presence of field nulls and closed field lines},
  \textit{Astrophysical Journal}, \textit{350}, 672--691, \doi{10.1086/168419}.

\bibitem[{\textit{{Lin} et~al.}(2005)\textit{{Lin}, {Ko}, {Sui}, {Raymond},
  {Stenborg}, {Jiang}, {Zhao}, and {Mancuso}}}]{Lin05}
{Lin}, J., Y.-K. {Ko}, L.~{Sui}, J.~C. {Raymond}, G.~A. {Stenborg}, Y.~{Jiang},
  S.~{Zhao}, and S.~{Mancuso} (2005), {Direct Observations of the Magnetic
  Reconnection Site of an Eruption on 2003 November 18}, \textit{\apj},
  \textit{622}, 1251--1264, \doi{10.1086/428110}.

\bibitem[{\textit{{Liu} et~al.}(2010)\textit{{Liu}, {Lee}, {Wang}, {Stenborg},
  {Liu}, and {Wang}}}]{Liu10}
{Liu}, R., J.~{Lee}, T.~{Wang}, G.~{Stenborg}, C.~{Liu}, and H.~{Wang} (2010),
  {A Reconnecting Current Sheet Imaged in a Solar Flare}, \textit{\apjl},
  \textit{723}, L28--L33, \doi{10.1088/2041-8205/723/1/L28}.

\bibitem[{\textit{{Longcope}}(2001)}]{Longcope01}
{Longcope}, D.~W. (2001), {Separator current sheets: Generic features in
  minimum-energy magnetic fields subject to flux constraints}, \textit{Physics
  of Plasmas}, \textit{8}, 5277--5290, \doi{10.1063/1.1418431}.

\bibitem[{\textit{{Longcope} and {Cowley}}(1996)}]{Longcope96}
{Longcope}, D.~W., and S.~C. {Cowley} (1996), {Current sheet formation along
  three-dimensional magnetic separators}, \textit{Physics of Plasmas},
  \textit{3}, 2885--2897, \doi{10.1063/1.871627}.

\bibitem[{\textit{{Longcope} and {Priest}}(2007)}]{Longcope07}
{Longcope}, D.~W., and E.~R. {Priest} (2007), {Fast magnetosonic waves launched
  by transient, current sheet reconnection}, \textit{Physics of Plasmas},
  \textit{14}(12), 122905, \doi{10.1063/1.2823023}.

\bibitem[{\textit{{Longcope} and {Tarr}}(2012)}]{LongcopeTarr12}
{Longcope}, D.~W., and L.~{Tarr} (2012), {The Role of Fast Magnetosonic Waves
  in the Release and Conversion via Reconnection of Energy Stored by a Current
  Sheet}, \textit{\apj}, \textit{756}, 192, \doi{10.1088/0004-637X/756/2/192}.

\bibitem[{\textit{{Milligan} et~al.}(2010)\textit{{Milligan}, {McAteer},
  {Dennis}, and {Young}}}]{Milligan10}
{Milligan}, R.~O., R.~T.~J. {McAteer}, B.~R. {Dennis}, and C.~A. {Young}
  (2010), {Evidence of a Plasmoid-Looptop Interaction and Magnetic Inflows
  During a Solar Flare/Coronal Mass Ejection Eruptive Event}, \textit{\apj},
  \textit{713}, 1292--1300, \doi{10.1088/0004-637X/713/2/1292}.

\bibitem[{\textit{{Moore} et~al.}(2002)\textit{{Moore}, {Fok}, and
  {Chandler}}}]{Moore02}
{Moore}, T.~E., M.-C. {Fok}, and M.~O. {Chandler} (2002), {The dayside
  reconnection X line}, \textit{Journal of Geophysical Research (Space
  Physics)}, \textit{107}, 1332, \doi{10.1029/2002JA009381}.

\bibitem[{\textit{{Nishizuka} et~al.}(2010)\textit{{Nishizuka}, {Takasaki},
  {Asai}, and {Shibata}}}]{Nishizuka10}
{Nishizuka}, N., H.~{Takasaki}, A.~{Asai}, and K.~{Shibata} (2010), {Multiple
  Plasmoid Ejections and Associated Hard X-ray Bursts in the 2000 November 24
  Flare}, \textit{\apj}, \textit{711}, 1062--1072,
  \doi{10.1088/0004-637X/711/2/1062}.

\bibitem[{\textit{{{\O}ieroset} et~al.}(2000)\textit{{{\O}ieroset}, {Phan},
  {Lin}, and {Sonnerup}}}]{Oieroset00}
{{\O}ieroset}, M., T.~D. {Phan}, R.~P. {Lin}, and B.~U.~{\"O}. {Sonnerup}
  (2000), {Wal{\'e}n and variance analyses of high-speed flows observed by Wind
  in the midtail plasma sheet: Evidence for reconnection}, \textit{\jgr},
  \textit{105}, 25,247--25,264, \doi{10.1029/2000JA900075}.

\bibitem[{\textit{{Parker}}(1957)}]{Parker57}
{Parker}, E.~N. (1957), {Sweet's Mechanism for Merging Magnetic Fields in
  Conducting Fluids}, \textit{Journal of Geophysical Research}, \textit{62},
  509--520, \doi{10.1029/JZ062i004p00509}.

\bibitem[{\textit{{Parnell} et~al.}(2008)\textit{{Parnell}, {Haynes}, and
  {Galsgaard}}}]{Parnell08}
{Parnell}, C.~E., A.~L. {Haynes}, and K.~{Galsgaard} (2008), {Recursive
  Reconnection and Magnetic Skeletons}, \textit{Astrophysical Journal},
  \textit{675}, 1656--1665, \doi{10.1086/527532}.

\bibitem[{\textit{{Parnell} et~al.}(2010)\textit{{Parnell}, {Haynes}, and
  {Galsgaard}}}]{Parnell10a}
{Parnell}, C.~E., A.~L. {Haynes}, and K.~{Galsgaard} (2010), {Structure of
  magnetic separators and separator reconnection}, \textit{Journal of
  Geophysical Research (Space Physics)}, \textit{115}, A02102,
  \doi{10.1029/2009JA014557}.

\bibitem[{\textit{{Parnell} et~al.}(2015)\textit{{Parnell}, {Stevenson},
  {Threlfall}, and {Edwards}}}]{Parnell15philtrans}
{Parnell}, C.~E., J.~E.~H. {Stevenson}, J.~{Threlfall}, and S.~J. {Edwards}
  (2015), {Is magnetic topology important for heating the solar atmosphere?},
  \textit{Philosophical Transactions of the Royal Society of London Series A},
  \textit{373}, 40,264, \doi{10.1098/rsta.2014.0264}.

\bibitem[{\textit{{Paschmann} et~al.}(1979)\textit{{Paschmann},
  {Papamastorakis}, {Sckopke}, {Haerendel}, {Sonnerup}, {Bame}, {Asbridge},
  {Gosling}, {Russel}, and {Elphic}}}]{Paschmann79}
{Paschmann}, G., I.~{Papamastorakis}, N.~{Sckopke}, G.~{Haerendel}, B.~U.~O.
  {Sonnerup}, S.~J. {Bame}, J.~R. {Asbridge}, J.~T. {Gosling}, C.~T. {Russel},
  and R.~C. {Elphic} (1979), {Plasma acceleration at the earth's magnetopause -
  Evidence for reconnection}, \textit{Nature}, \textit{282}, 243--246,
  \doi{10.1038/282243a0}.

\bibitem[{\textit{{Petschek}}(1964)}]{Petschek64}
{Petschek}, H.~E. (1964), {Magnetic Field Annihilation}, \textit{NASA Special
  Publication}, \textit{50}, 425.

\bibitem[{\textit{{Phan} et~al.}(2000)\textit{{Phan}, {Kistler}, {Klecker},
  {Haerendel}, {Paschmann}, {Sonnerup}, {Baumjohann}, {Bavassano-Cattaneo},
  {Carlson}, {DiLellis}, {Fornacon}, {Frank}, {Fujimoto}, {Georgescu},
  {Kokubun}, {Moebius}, {Mukai}, {{\O}ieroset}, {Paterson}, and
  {Reme}}}]{Phan00}
{Phan}, T.~D., L.~M. {Kistler}, B.~{Klecker}, G.~{Haerendel}, G.~{Paschmann},
  B.~U.~{\"O}. {Sonnerup}, W.~{Baumjohann}, M.~B. {Bavassano-Cattaneo}, C.~W.
  {Carlson}, A.~M. {DiLellis}, K.-H. {Fornacon}, L.~A. {Frank}, M.~{Fujimoto},
  E.~{Georgescu}, S.~{Kokubun}, E.~{Moebius}, T.~{Mukai}, M.~{{\O}ieroset},
  W.~R. {Paterson}, and H.~{Reme} (2000), {Extended magnetic reconnection at
  the Earth's magnetopause from detection of bi-directional jets},
  \textit{Nature}, \textit{404}, 848--850.

\bibitem[{\textit{{Platten} et~al.}(2014)\textit{{Platten}, {Parnell},
  {Haynes}, {Priest}, and {Mackay}}}]{Platten14}
{Platten}, S.~J., C.~E. {Parnell}, A.~L. {Haynes}, E.~R. {Priest}, and D.~H.
  {Mackay} (2014), {The solar cycle variation of topological structures in the
  global solar corona}, \textit{\aap}, \textit{565}, A44,
  \doi{10.1051/0004-6361/201323048}.

\bibitem[{\textit{{Pontin} and {Craig}}(2006)}]{PC06}
{Pontin}, D.~I., and I.~J.~D. {Craig} (2006), {Dynamic Three-dimensional
  Reconnection in a Separator Geometry with Two Null Points}, \textit{\apj},
  \textit{642}, 568--578, \doi{10.1086/500725}.

\bibitem[{\textit{{Priest} and {Forbes}}(2000)}]{PriestForbes}
{Priest}, E., and T.~{Forbes} (Eds.) (2000), \textit{{Magnetic reconnection :
  MHD theory and applications}}.

\bibitem[{\textit{{Priest} et~al.}(2005)\textit{{Priest}, {Longcope}, and
  {Heyvaerts}}}]{Priest05}
{Priest}, E.~R., D.~W. {Longcope}, and J.~{Heyvaerts} (2005), {Coronal Heating
  at Separators and Separatrices}, \textit{\apj}, \textit{624}, 1057--1071,
  \doi{10.1086/429312}.

\bibitem[{\textit{{R{\`e}me} et~al.}(2001)\textit{{R{\`e}me}, {Aoustin},
  {Bosqued}, {Dandouras}, {Lavraud}, {Sauvaud}, {Barthe}, {Bouyssou}, {Camus},
  {Coeur-Joly}, {Cros}, {Cuvilo}, {Ducay}, {Garbarowitz}, {Medale}, {Penou},
  {Perrier}, {Romefort}, {Rouzaud}, {Vallat}, {Alcayd{\'e}}, {Jacquey},
  {Mazelle}, {D'Uston}, {M{\"o}bius}, {Kistler}, {Crocker}, {Granoff},
  {Mouikis}, {Popecki}, {Vosbury}, {Klecker}, {Hovestadt}, {Kucharek},
  {Kuenneth}, {Paschmann}, {Scholer}, {Sckopke}, {Seidenschwang}, {Carlson},
  {Curtis}, {Ingraham}, {Lin}, {McFadden}, {Parks}, {Phan}, {Formisano},
  {Amata}, {Bavassano-Cattaneo}, {Baldetti}, {Bruno}, {Chionchio}, {di Lellis},
  {Marcucci}, {Pallocchia}, {Korth}, {Daly}, {Graeve}, {Rosenbauer},
  {Vasyliunas}, {McCarthy}, {Wilber}, {Eliasson}, {Lundin}, {Olsen}, {Shelley},
  {Fuselier}, {Ghielmetti}, {Lennartsson}, {Escoubet}, {Balsiger}, {Friedel},
  {Cao}, {Kovrazhkin}, {Papamastorakis}, {Pellat}, {Scudder}, and
  {Sonnerup}}}]{Reme01}
{R{\`e}me}, H., C.~{Aoustin}, J.~M. {Bosqued}, I.~{Dandouras}, B.~{Lavraud},
  J.~A. {Sauvaud}, A.~{Barthe}, J.~{Bouyssou}, T.~{Camus}, O.~{Coeur-Joly},
  A.~{Cros}, J.~{Cuvilo}, F.~{Ducay}, Y.~{Garbarowitz}, J.~L. {Medale},
  E.~{Penou}, H.~{Perrier}, D.~{Romefort}, J.~{Rouzaud}, C.~{Vallat},
  D.~{Alcayd{\'e}}, C.~{Jacquey}, C.~{Mazelle}, C.~{D'Uston}, E.~{M{\"o}bius},
  L.~M. {Kistler}, K.~{Crocker}, M.~{Granoff}, C.~{Mouikis}, M.~{Popecki},
  M.~{Vosbury}, B.~{Klecker}, D.~{Hovestadt}, H.~{Kucharek}, E.~{Kuenneth},
  G.~{Paschmann}, M.~{Scholer}, N.~{Sckopke}, E.~{Seidenschwang}, C.~W.
  {Carlson}, D.~W. {Curtis}, C.~{Ingraham}, R.~P. {Lin}, J.~P. {McFadden},
  G.~K. {Parks}, T.~{Phan}, V.~{Formisano}, E.~{Amata}, M.~B.
  {Bavassano-Cattaneo}, P.~{Baldetti}, R.~{Bruno}, G.~{Chionchio}, A.~{di
  Lellis}, M.~F. {Marcucci}, G.~{Pallocchia}, A.~{Korth}, P.~W. {Daly},
  B.~{Graeve}, H.~{Rosenbauer}, V.~{Vasyliunas}, M.~{McCarthy}, M.~{Wilber},
  L.~{Eliasson}, R.~{Lundin}, S.~{Olsen}, E.~G. {Shelley}, S.~{Fuselier}, A.~G.
  {Ghielmetti}, W.~{Lennartsson}, C.~P. {Escoubet}, H.~{Balsiger},
  R.~{Friedel}, J.-B. {Cao}, R.~A. {Kovrazhkin}, I.~{Papamastorakis},
  R.~{Pellat}, J.~{Scudder}, and B.~{Sonnerup} (2001), {First multispacecraft
  ion measurements in and near the Earth's magnetosphere with the identical
  Cluster ion spectrometry (CIS) experiment}, \textit{Annales Geophysicae},
  \textit{19}, 1303--1354, \doi{10.5194/angeo-19-1303-2001}.

\bibitem[{\textit{{Savage} et~al.}(2012)\textit{{Savage}, {Holman}, {Reeves},
  {Seaton}, {McKenzie}, and {Su}}}]{Savage12}
{Savage}, S.~L., G.~{Holman}, K.~K. {Reeves}, D.~B. {Seaton}, D.~E. {McKenzie},
  and Y.~{Su} (2012), {Low-altitude Reconnection Inflow-Outflow Observations
  during a 2010 November 3 Solar Eruption}, \textit{\apj}, \textit{754}, 13,
  \doi{10.1088/0004-637X/754/1/13}.

\bibitem[{\textit{{Schindler} et~al.}(1988)\textit{{Schindler}, {Hesse}, and
  {Birn}}}]{Schindler88}
{Schindler}, K., M.~{Hesse}, and J.~{Birn} (1988), {General magnetic
  reconnection, parallel electric fields, and helicity}, \textit{Journal of
  Geophysical Research}, \textit{93}, 5547--5557,
  \doi{10.1029/JA093iA06p05547}.

\bibitem[{\textit{{Sonnerup}}(1974)}]{Sonnerup74}
{Sonnerup}, B.~U.~{\"O}. (1974), {Magnetopause reconnection rate},
  \textit{\jgr}, \textit{79}, 1546--1549, \doi{10.1029/JA079i010p01546}.

\bibitem[{\textit{{Sonnerup}}(1979)}]{Sonnerup79}
{Sonnerup}, B.~U.~{\"O}. (1979), \textit{{Magnetic field reconnection}}, pp.
  45--108.

\bibitem[{\textit{{Sonnerup} et~al.}(1981)\textit{{Sonnerup}, {Paschmann},
  {Papamastorakis}, {Sckopke}, {Haerendel}, {Bame}, {Asbridge}, {Gosling}, and
  {Russell}}}]{Sonnerup81}
{Sonnerup}, B.~U.~O., G.~{Paschmann}, I.~{Papamastorakis}, N.~{Sckopke},
  G.~{Haerendel}, S.~J. {Bame}, J.~R. {Asbridge}, J.~T. {Gosling}, and C.~T.
  {Russell} (1981), {Evidence for magnetic field reconnection at the earth's
  magnetopause}, \textit{\jgr}, \textit{86}, 10,049--10,067,
  \doi{10.1029/JA086iA12p10049}.

\bibitem[{\textit{{Stevenson} and {Parnell}}(2015)}]{Stevenson15_jgra}
{Stevenson}, J.~E.~H., and C.~E. {Parnell} (2015), {Spontaneous reconnection at
  a separator current layer. I. Nature of the reconnection}, \textit{Journal of
  Geophysical Research}.

\bibitem[{\textit{{Stevenson} et~al.}(2015)\textit{{Stevenson}, {Parnell},
  {Priest}, and {Haynes}}}]{Stevenson15}
{Stevenson}, J.~E.~H., C.~E. {Parnell}, E.~R. {Priest}, and A.~L. {Haynes}
  (2015), {The nature of separator current layers in MHS equilibria. I. Current
  parallel to the separator}, \textit{\aap}, \textit{573}, A44,
  \doi{10.1051/0004-6361/201424348}.

\bibitem[{\textit{{{Stevenson}, J. E. H.}}(2015)}]{Stevenson15PhD}
{{Stevenson}, J. E. H.} (2015), On the properties of single-separator mhs
  equilibria and the nature of separator reconnection, Ph.D. thesis, School of
  Mathematics and Statistics, University of St Andrews.

\bibitem[{\textit{{Swisdak} and {Drake}}(2007)}]{Swisdak07}
{Swisdak}, M., and J.~F. {Drake} (2007), {Orientation of the reconnection
  X-line}, \textit{\grl}, \textit{34}, L11106, \doi{10.1029/2007GL029815}.

\bibitem[{\textit{{Takasao} et~al.}(2012)\textit{{Takasao}, {Asai}, {Isobe},
  and {Shibata}}}]{Takasao12}
{Takasao}, S., A.~{Asai}, H.~{Isobe}, and K.~{Shibata} (2012), {Simultaneous
  Observation of Reconnection Inflow and Outflow Associated with the 2010
  August 18 Solar Flare}, \textit{\apjl}, \textit{745}, L6,
  \doi{10.1088/2041-8205/745/1/L6}.

\bibitem[{\textit{{Trattner} et~al.}(2007)\textit{{Trattner}, {Mulcock},
  {Petrinec}, and {Fuselier}}}]{Trattner07}
{Trattner}, K.~J., J.~S. {Mulcock}, S.~M. {Petrinec}, and S.~A. {Fuselier}
  (2007), {Probing the boundary between antiparallel and component reconnection
  during southward interplanetary magnetic field conditions}, \textit{Journal
  of Geophysical Research (Space Physics)}, \textit{112}, A08210,
  \doi{10.1029/2007JA012270}.

\bibitem[{\textit{{Trenchi} et~al.}(2008)\textit{{Trenchi}, {Marcucci},
  {Pallocchia}, {Consolini}, {Bavassano Cattaneo}, {di Lellis}, {R{\`e}Me},
  {Kistler}, {Carr}, and {Cao}}}]{Trenchi2008}
{Trenchi}, L., M.~F. {Marcucci}, G.~{Pallocchia}, G.~{Consolini}, M.~B.
  {Bavassano Cattaneo}, A.~M. {di Lellis}, H.~{R{\`e}Me}, L.~{Kistler}, C.~M.
  {Carr}, and J.~B. {Cao} (2008), {Occurrence of reconnection jets at the
  dayside magnetopause: Double Star observations}, \textit{Journal of
  Geophysical Research (Space Physics)}, \textit{113}, A07S10,
  \doi{10.1029/2007JA012774}.

\bibitem[{\textit{{Wang} et~al.}(2007)\textit{{Wang}, {Sui}, and
  {Qiu}}}]{Wang07}
{Wang}, T., L.~{Sui}, and J.~{Qiu} (2007), {Direct Observation of High-Speed
  Plasma Outflows Produced by Magnetic Reconnection in Solar Impulsive Events},
  \textit{\apjl}, \textit{661}, L207--L210, \doi{10.1086/519004}.

\bibitem[{\textit{{Watanabe} et~al.}(2012)\textit{{Watanabe}, {Hara},
  {Sterling}, and {Harra}}}]{Watanabe12}
{Watanabe}, T., H.~{Hara}, A.~C. {Sterling}, and L.~K. {Harra} (2012),
  {Production of High-Temperature Plasmas During the Early Phases of a C9.7
  Flare. II. Bi-directional Flows Suggestive of Reconnection in a Pre-flare
  Brightening Region}, \textit{Solar Physics}, \textit{281}, 87--99,
  \doi{10.1007/s11207-012-0079-5}.

\bibitem[{\textit{{Yokoyama} et~al.}(2001)\textit{{Yokoyama}, {Akita},
  {Morimoto}, {Inoue}, and {Newmark}}}]{Yokoyama01}
{Yokoyama}, T., K.~{Akita}, T.~{Morimoto}, K.~{Inoue}, and J.~{Newmark} (2001),
  {Clear Evidence of Reconnection Inflow of a Solar Flare}, \textit{\apjl},
  \textit{546}, L69--L72, \doi{10.1086/318053}.

\end{thebibliography}

\end{article}
\end{document}